\documentclass{JHEP3}
\usepackage{graphicx}
\usepackage{amssymb}

\def\beq{\begin{equation}}
\def\enq{\end{equation}}
\def\bea{\begin{eqnarray}}
\def\ena{\end{eqnarray}}
\newcommand{\non}{\nonumber \\}
\def\pa{\partial}
\newcommand{\rz}{\{r,z\}}
\newcommand{\rhoxi}{\{\rho,\xi\}}
\newcommand{\IS}{{\bf S}}
\def\sqrtmu{\sqrt{1-\xi^2}}
\def\sqmu{1-\xi^2}
\def\det{\sqrt{-g}}
\def\G{{\cal G}}
\def\Sch{Schwarzschild}

\def\bhh{S^3}
 \def\eps{\epsilon}
 \def\hL{\hat{L}}
\def\hA{\hat{A}}
\def\hB{\hat{B}}
\def\hC{\hat{C}}

\newcommand{\sbsection}[1]{\vspace{.5cm} \noindent {\it #1}}
\preprint{{\tt hep-th/0310096}}

\title{ \center{}Caged Black Holes:  \\ Black Holes in
Compactified Spacetimes II  --  \\ 5d Numerical Implementation}

\author{Evgeny Sorkin, Barak Kol, Tsvi Piran
\\
 Racah Institute of Physics\\
 Hebrew University \\
 Jerusalem 91904,
 Israel\\
{\tt sorkin, barak\_kol,  tsvi @phys.huji.ac.il}}

\abstract{ We describe the first convergent numerical method to
  determine static black hole solutions (with $S^3$ horizon) in 5d
  compactified spacetime. We obtain a family of solutions parametrized
  by the ratio of the black hole size and the size of the compact
  extra dimension. The solutions satisfy the demanding integrated
  first law. For small black holes our solutions approach the 5d
  Schwarzschild solution and agree very well with new theoretical
  predictions for the small corrections to thermodynamics and
  geometry. The existence of such black holes is thus established.  We
  report on thermodynamical (temperature, entropy, mass and tension
  along the compact dimension) and geometrical measurements. Most
  interestingly, for large masses (close to the Gregory-Laflamme
  critical mass) the scheme destabilizes. We interpret this as evidence
  for an approach to a physical tachyonic instability. Using
  extrapolation we speculate that the system undergoes a
  first order phase transition.}

\begin{document}

\section{Introduction}
\label{intro}
In backgrounds with additional compact dimensions there may exist
several phases of black objects including black-holes and
black-strings. The phase transition between these phases raises
puzzles and touches fundamental issues such as topology change,
uniqueness and Cosmic Censorship.

Consider for concreteness a background with a single compact dimension
-- ${\mathbb R}^{d-2,1} \times \IS^1$. We denote the coordinate along
the compact dimension by $z$ and the period by $\hL$.  The problem is
characterized by a single dimensionless parameter\footnote{ Later we
  will use another parameter $x$ defined in (\ref{iks}).}, e.g.  the
dimensionless mass, $\mu=G_N M/\hL^{d-3}$ where $G_N$ is the $d$
dimensional Newton constant and $M$ is the (asymptotic) mass. Gregory
and Laflamme (GL) \cite{GL1,GL2} discovered that a uniform black
string -- the $d-1$ \Sch \ solution times a line -- becomes
classically unstable below a certain  critical value $\mu_{\rm GL}$.
They interpreted this instability as a decay of the string to a single
localized black hole.  Their discovery has initiated intensive
research\footnote{Related research includes
  \cite{MyersPerry,Myers,wis3,KudohTanakaNakamura,EmparanMyers}.}
\cite{HM1,HM2,gubser,deSmet,HO1,barak1,wis1,wis2,barak2,barak3,KolWiseman,sorpir,dynamics,HO2,HO3}
that attempted to trace out the fate of the unstable GL
string: whether it settles at another intermediate stable phase as
advocated in \cite{HM1,HM2}, or whether it really decays to a single black
hole. By now there is  mounting direct evidence against the former
possibility \cite{barak1,wis1}, together with  additional
circumstantial evidence \cite{gubser,deSmet}; and  \cite{dynamics}
which we also regard as evidence against stable non-uniform black
string phase\footnote{Note however that the authors of \cite{dynamics}  did not
interpret their results either as supporting or as countering the
conjecture.}.

Here motivated by \cite{barak1} we take another route, namely we
address the question: what happens to a small localized black hole as
its mass increases (by e.g. absorption of an interstellar dust)? Such
a black hole grows and naively one expects that there is a moment when
its ``north'' and ``south'' poles touch. Whether this is the case or
not is yet to be established, but it is clear that some sort of
instability will show up when the poles are getting closer. Put
differently: is there a maximal mass, beyond which the black hole
``does not fit into the circle'' and  there
  are no stable black holes.  This maximal mass
  would be analogous to the GL critical mass, and would correspond to
  a perturbative, tachyonic instability . Yet another kind of
instability may occur before that maximal mass is reached. Once the
entropy of a black hole equals the entropy of uniform black string
{\it with the same mass}, a transition between both phases will be
allowed by quantum tunneling, or by thermal fluctuations. This first
order phase transition is slower than the classical perturbative
instability due to tunneling suppression.

No analytical solution for a black hole is known. Moreover, even
though one can expect approximate analytic solutions to exist for very
small black holes, the phase transition physics happens when the size
of the black hole is comparable to the size of the compact dimension.
Hence, in this work we take the numerical avenue.

In our first article \cite{numericI} we considered the theoretical
background for the static d-dimensional quasi-spherical black holes
(BHs). There we outlined the goals of the numerical study. Prime among
these goals is to establish the very existence of the static black
hole solutions. To our knowledge, there is no direct evidence in the
literature that such BHs do exist\footnote{Arguments like: ``in the
  limit when the radius of a BH is small compared to the
  compactification radius the equivalence principle implies that the black hole must be
  similar to the 5d \Sch \ solution'', while intuitive are not
  rigorously sufficient. In fact, this argument fails in 4d with one
  of the space-like directions being curled to a circle, as there is no
  stable configuration of a periodic array of point-like sources.}
though there are positive indications for that \cite{sorpir}. Among
other goals is the study of such BH solutions in various regimes and
dimensions. The ultimate and the most interesting aim is of course to
determine the point of phase transition.

Based on recent progress \cite{wis1,KudohTanakaNakamura} we develop a
numerical scheme that allows us to find static axisymmetric BH
solutions.  Our scheme is dimension independent, provided that $d>4$.
As a first step, in this paper, we apply it to the 5d case, which is
the example with the lowest dimension\footnote{This is maybe the
  lowest dimensional example, but because of a very slow asymptotic
  decay, it is certainly not the simplest to solve numerically
  \cite{gubser}. We discuss this in detail later on.} among spacetimes
with extra dimensions.  In 5d we construct numerically a family of
static BHs, parametrized by $x$, which is the ratio of the size of the
black hole to the size of the compact dimension, see (\ref{iks}).  For
small values of $x$ the horizon region of our solutions approaches the
5d \Sch \ solution which can be considered as the 'zeroth order in a
perturbative expansion' in powers of $x$.  Moreover, in this limit our
solutions satisfy the theoretical expectations for some next order
corrections in this perturbative analysis \cite{GorbonosKol} thereby
allowing a confirmation of a new theoretical method through numerical
``experiment''. This establishes the existence of static
higher-dimensional BHs and shows that the \Sch \ solution is the
smooth limit of these solutions for $x \rightarrow 0$.

We succeed to control the accuracy of our solutions up to $x \lesssim
x_1\simeq0.20$ (corresponding to $\mu_1\simeq0.047$). Above this
limit, up to the last value $x_2\simeq0.25$ (corresponding to
$\mu_2\simeq0.074$), for which our solutions do not diverge the
convergence rate was very slow and the numerical errors were not
small. These values of $\mu$ should be compared with the critical GL
mass $\mu_{\rm GL} \simeq 0.070$. The slowdown of convergence and
eventual divergence is mainly seen on one of our metric functions. By
examining the equations of motion for our system we observe a "wrong"
sign in one of the equations (just like the plus sign in the following
harmonic oscillator equation, $\psi'' +\omega_o^2 \psi =0$), which is
an indication of the presence of the tachyon.  One could expect that
the tachyonic behavior is suppressed for small $x$ values and it is
manifest for large $x$ values, for which there are no static BH solutions\footnote{ Consider a tachyon in a box: the mode  can
  materialize only if its inverse mass is not less then the dimension of
  the box otherwise the mode is suppressed.} 
\footnote{ This is a classical ``revolutionary
  situation'': a ``poor'' tachyon is suppressed until the black hole
  becomes too fat.  Then the tachyon rises, gets strong and destroys
  the black hole, heading to a new future (to another
  phase).}.
However, the tachyonic behavior influences the numerics even before
that critical $x$ value and slows down the convergence.  We believe
that the problematic variable is coupled to the tachyonic mode, and
hence when the latter drives the former to behave pathologically, is
an indication that the system is  close to the
  phase transition point.

In \cite{numericI} we derived the d-dimensional Smarr formula, also
known as the integrated first law, for the geometry under study (see
also \cite{HO2}). It is a relation between thermodynamic quantities at
the horizon and those at infinity, relying on the generalized Stokes
formula and the validity of the equations of motion in the interior.
This naturally suggests to use this formula to estimate the ``overall
numerical error'' in our numerical implementations. This method comes
in addition to the standard numerical tests such as convergence rate,
constraint violation etc.  While it is possible that this is a lucky
coincidence (though we believe it is not), for our solutions the Smarr
formula is satisfied with $3-4\%$ accuracy. Moreover, it is intriguing enough
that the Smarr formula is satisfied with the same $4\%$ accuracy even
for the problematic solutions for $x \gtrsim 0.20$. This has to do
with the fact that this formula relates only 3 variables of the 4
thermodynamical variables, characterizing the system.  It turns out
that the 4th variable is somewhat decoupled from the other three.
However, as the inaccuracy in determining this 4th variable grows with
$x$, this slows down and ultimately ruins the convergence.  We believe
that this is the variable which is coupled to the tachyonic mode.

Even though the Smarr formula is satisfied to a good accuracy, we do
have some larger inaccuracies in the solutions. One of the fields
suffers from a certain convergence problems, and its asymptotic
behavior departs by some $30 \%$ from small $x$ predictions. This is
exactly the ``fourth'' asymptotic charge that does not appear in
Smarr's formula. Due to its approximate decoupling it {\bf is }plausible that
indeed we have good accuracy for all other measurements. Even better,
we have indications that this field is reliable for a sub-range of
$x$: $0.08 \lesssim x \lesssim 0.15$.

 One can question what
information can be extracted from knowledge of only three parameters
for the entire sequence of BHs ($x \lesssim 0.25$), and knowledge of
the fourth one for a smaller range ($0.08 \lesssim x \lesssim 0.15$).
In particular, we show that the last black hole that we find (at
$x\simeq0.25$) deviates only slightly from being spherical, and moreover,
its poles are quite distant from each other.

In addition, one can ask whether there is a first order phase
transition.  We cannot establish this with certainty, since the
entropy of our last black hole (at $x_2\simeq 0.25$) is still larger
than the entropy of the corresponding uniform black string. A naive
extrapolation of our data to larger values of $x$ indicates that the
entropies will become equal just above the maximal BH that we find,
namely at $x_3\simeq 0.26$ that corresponds to $\mu_3 \simeq 0.082$.
It is rather suggestive that $\mu_{GL},\mu_2$ and $\mu_3$ are all very
close each to another.  Since, all the numbers in the system are
expected to be of the same order, this fact may be regarded as an
indication that we have found a real phase transition. Note that since
$\mu_2\simeq\mu_{\rm GL}$ we come close to a first demonstration of a
failure of higher dimensional uniqueness with two {\it stable}
phases\footnote{Although we did not demonstrate that we assume that
  our BHs solutions are stable. }. Finally we note, that generally
in a first order phase transition one expects $\mu_{\rm GL}\leq
\mu_3\leq\mu_2$. This remains to be tested numerically.

While we expect that the instability we found corresponds to a physical
one we stress that we cannot rule out the conservative possibility
that it is a manifestation of imperfections of the numerics.  Since
our numerical scheme is independent of the dimensionality of the
problem provided $d>4$, the immediate aim for the future work would be
its application to higher dimensions, $d \ge 6$, where the asymptotic fall
off is faster and the solutions might be more stable\footnote{In
  fact the preliminary results show that the picture
  that we find in 5d is qualitatively unchanged for  $d \ge 6$.}.

In section \ref{sec_equations} we describe our system. We employ the
``conformal ansatz'' and derive the equations of motion and the boundary
conditions. A short excursion into theoretical background (summarized
from \cite{numericI}) is made in section \ref{sec_theory}.  Our
numerical implementation is described in detail in section
\ref{sec_numerics}, where we also describe various tests. The results
are listed in section \ref{sec_results}. We outline future directions
in the final section \ref{sec_discussion}.  In appendix
\ref{Equations_appendix}, we derive the d-dimensional field equations
and boundary conditions for the cylinder ${\mathbb R}^{d-2,1} \times
\IS^1$. In appendix \ref{appendix_asympt} we consider the asymptotic
behavior of the equations.

We also refer the reader to independent work by Kudoh and Wiseman who
performed recently related calculations in 6d \cite{KudohWiseman}. 
\section{Formulation}
\label{sec_equations}
In this section we focus on the five dimensional case -- we derive
the field equations and discuss the boundary conditions (b.c.).
Equations and b.c. on a general d-cylinder are discussed in
appendix \ref{Equations_appendix} . The fifth spatial direction is
denoted by $z$ and it is compact with a period $\hL$, i.e.  $z$
and $z+\hL$ are identified . We consider static localized BHs with an $\bhh$ horizon
topology. We assume spherical symmetry ($SO(3)$ isometry) of the 3
extended spatial dimensions and we denote the 4d radial
coordinate by $r$.
\subsection{Choice of coordinates}
\label{sec_ansatz}
We consider a static axisymmetric metric which is built out of three
functions.  We adopt a conformal (in the $\{r,z\}$ plane) ansatz of
the form
\beq
\label{metric_rz}
ds^2=-A^2 dt^2+e^{2B}(dr^2+dz^2)+e^{2C} r^2 d\Omega^2_2 \ ,
\enq
where $A,B$ and $C$ are functions of $r,z$ only and
$d\Omega^2_2= d\theta^2 +\sin^2\theta\, d\phi^2$

To describe a BH it is convenient to transform to polar
coordinates, defined by
\beq
\label{polarcoord}
r=\rho \sin \chi, ~~ \ z=\rho \cos \chi,
\enq
 since the BH horizon is represented by a closed curve in
the $\{r,z\}$ plane.
The metric  in these coordinates reads now
\beq
\label{metric_Rchi}
ds^2=-A^2 dt^2+e^{2B}(d\rho^2+\rho^2d\chi^2)+e^{2C} \rho^2\sin^2\chi
d\Omega^2_2 \ ,
\enq

\begin{figure}[t!]
\centering
\noindent
\includegraphics[width=14cm]{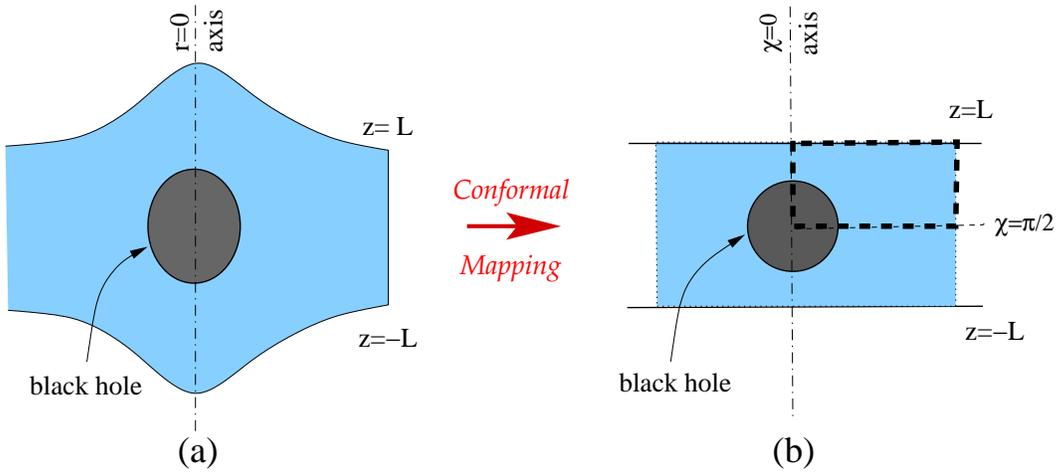}
\caption[]{A spacelike slice of
  the black-hole spacetime. $(a)$ In the $\rz$ plane the black
  hole's horizon is a curve with a spherical $\bhh$ topology. $(b)$
  There is a conformal freedom to transform the domain to $\{(r,z): |z|
  \le L, ~r^2+z^2 \ge \rho_h^{~2} \}$. By fixing $\rho_h/L$ the domain
  is uniquely specified \cite{KudohTanakaNakamura}. }
\label{fig_map}
\end{figure}

To simplify the numerical procedure it is desirable that the
boundaries of the integration domain\footnote{What we call here
``domain of integration"  could be called alternatively ``domain of
definition", ``domain of relaxation" etc. By this term we refer to
the region of space-time where we solve our equations.} 
lie along the coordinate lines. Note that by choosing the ansatz
(\ref{metric_rz}), or (\ref{metric_Rchi}) we still did not fix the
gauge completely. There is still a freedom to move the
boundaries of the integration domain by a conformal
transformation.  It was shown in \cite{KudohTanakaNakamura} that using this
conformal freedom the horizon boundary could be set at a constant
radius $\rho_h$, leaving the periodic boundaries
along $z=const$ lines.  Thus the domain is $\{(r,z): |z| \le L,
~r^2+z^2 \ge \rho_h^{~2} \}$, where for future use we define the
half-period $L=\hL/2$ of the compact circle, see Fig.
(\ref{fig_map}). In addition, by fixing the ratio of the radius of
the horizon to the period of the circle
 \beq
\label{iks}
x:= {\rho_h \over L}, \enq
all residual gauge freedom is eliminated. In our implementation,
we set  $\rho_h=1$, without a loss of generality, and generate
different solutions by varying $L$.

The fact that $x$ cannot be changed freely for a given solution
implies that $x$ is a characteristic parameter analogous to the
(normalized) total mass or the temperature even though it does not
have a clear physical meaning. For example, if one enforces the
horizon of a BH to be at a fixed radius set by some given $x$, it
would be excessive to specify also the temperature. Conversely,
specifying the temperature one does not have the  freedom to
constrain the location of the horizon \cite{KudohTanakaNakamura}.

In polar coordinates the reflecting  boundary of the compact
circle, $z=0$, is at  $\chi=\pi/2$, but the periodic  boundary,
$z=L$, does not lie along a coordinate line in the $\{\rho,\chi\}$
plane. The treatment of this irregular boundary introduces a
certain complication in the numerical scheme as described in
section (\ref{subsec_scheme}). Nevertheless, we believe that it is
preferable to work in polar coordinates (\ref{polarcoord}) and to
have an irregular boundary at $z=L$, rather than work in
rectangular coordinates $\{r,z\}$ and have an irregular boundary
at the horizon.  Intuitively, this is because we expect that the
region near the horizon would become the region of the 'activity'
as $ x$ increases.

For numerical reasons it would be convenient to use another angular
coordinate
\beq
\label{xi}
\xi=\cos(\chi) \ .
\enq
The benefit of using this coordinate is twofold. First, the
irregular $z=L$ boundary has a particularly simple representation,
$\rho=L/\xi$.  Second, as we explain shortly, the coordinate
singularity at the axis, $r=0$, becomes first order instead of
second order.
\subsection{Equations of motion}
\label{sec_EE}
Our basic equations are the five-dimensional time independent
vacuum Einstein equations.  There are five equations in 5d: two
are equations of motion for $A,C$, while variation with respect to
the metric in the $(r,z)$ plane yields three additional equations.
In the conformal ansatz one of them is an equation of motion for
$B$ while the other two result from the gauge fixing.  The
equations can be combined in a way that three of them will take
the form of elliptic equations, which we call the interior
equations. The other two combinations that contain a hyperbolic
differential operator will be termed 'the constraints'. These
constraint equations are not independent as they are related to
the interior equations via the Bianchi identities.

In order to obtain the interior equations we can follow the
general procedure described in appendix \ref{Equations_appendix},
or alternatively, use a suitable symbolic math application e.g.
GRTensor \cite{grii} to evaluate the relevant quantities. In
either route one obtains the interior equations which are the
following combinations of the components of the Einstein tensor:
$\G^\theta_\theta +{1/2} \G_\chi^\chi +{1/2}\G^\rho_\rho -2\G_t^t
\ , 2 \G^\theta_\theta -{2} \G_\chi^\chi -{2}\G^\rho_\rho +\G_t^t
$ and $\G^\theta_\theta + \G_\chi^\chi +\G^\rho_\rho -\G_t^t $.
They can be written respectively as
\begin{eqnarray}
  \label{EqA}
  \triangle A &+&
  {2\pa_\xi A \over \rho^2}\left(-\xi+(\sqmu) \pa_\xi C\right)
  +2 \pa_\rho A \left({1\over \rho} +\pa_\rho C\right)=0,  \\
  \label{EqB}
  \triangle B  &+&
  \frac{2\,\pa_\xi A\,\left(\xi-\left(1-{\xi }^2\right)\,\pa_\xi C \right) }{A\,\rho^2}+
  \frac{2\, \pa_\xi
    C\,\left(\xi-{1\over2}\left(1-{\xi}^2\right)\,\pa_\xi C \right)
    }{\rho^2} -
  \nonumber\\
  &-& \frac{2\,\pa_\rho C}{\rho} - \left({\pa_\rho C}\right)^2 -
  \frac{2\,\pa_\rho A}{A}\left( \frac{1}{\rho} + \pa_\rho C \right)-
  \frac{1 - e^{2\,B - 2\,C}}{\rho^2\,\left( 1 - {\xi }^2 \right) } =0\ , \\
  \label{EqC}
  \triangle C &-&
  \frac{\pa_\xi A\, \left( \xi-\left(1 -{\xi }^2 \right) \,\pa_\xi C \right) }{A\,\rho^2} -
  \frac{4\,\pa_\xi
    C\,\left(\xi-{1\over2}\left(1-{\xi}^2\right)\,\pa_\xi
      C\right)}{\rho^2} +
  \nonumber\\
  &+& \frac{4\,\pa_\rho C}{\rho} + 2\,\left({\pa_\rho C}\right)^2 +
  \frac{\pa_\rho A}{A}\,\left( \frac{1}{\rho} + \pa_\rho C \right) +
\frac{1 - e^{2\,\left( B - C \right) }}{\rho^2\,\left( 1 - {\xi
}^2 \right) }=0 \ .
\end{eqnarray}
Here we used the variable $\xi$ instead of $\chi$ and
 the Laplacian becomes
$\triangle \equiv \pa_\rho^2+(1/\rho)
\partial_\rho+(1/\rho^2)\sqrtmu\pa_\xi\left(\sqrtmu \pa_\xi\right)$.

The constraint equations expand to
\begin{eqnarray}
  \label{GRchi}
  \G_\rho^\xi &\equiv&
  \frac{{\sqrt{1-{\xi}^2}}}{\rho^2} \Big\{\frac{\pa_\xi A\,\left(\frac{1}{\rho}+\pa_\rho B\right)}{A}-\frac{2\,\xi \,\left( \pa_\rho B - \pa_\rho C \right)}{1-{\xi}^2}
+\frac{2\,\pa_\xi B}{\rho}+ 2\,\pa_\xi C\,\pa_\rho B
\nonumber\\
 &+& 2\,\pa_\xi B\,\pa_\rho C -
  2\,\pa_\xi C\,\pa_\rho C +
  \frac{\pa_\xi B\,\pa_\rho A - \pa_{\rho\xi} A }{A}
- 2\,\pa_{\rho\xi} C \Big\} =0\, \\
\label{GRR-Gchichi} \G_\rho^\rho-\G_\xi^\xi&\equiv & -\frac{\pa_\xi
A\,\left(\xi -2\,\left(1-{\xi }^2 \right) \,\pa_\xi B \right)
}{A\,\rho^2} + \frac{4\,\xi \,\left( \pa_\xi B - \pa_\xi C \right)
}{\rho^2}-
 \nonumber\\
&-&\frac{2\,\left(1-{\xi}^2\right)\,\left(2\,\pa_\xi B-\pa_\xi
C\right)\,\pa_\xi C}{\rho^2} +\frac{\left(1 - {\xi }^2 \right)
\,\pa_\xi^2 A}{A\,\rho^2}+
 \nonumber\\
&+& \frac{2\,\left( - \xi \,\pa_\xi C   + \left(1 - {\xi }^2
\right) \,\pa_\xi^2 C \right)}{\rho^2} + \frac{\pa_\rho
A\,\left( \frac{1}{\rho} + 2\,\pa_\rho B \right) }{A}+
\nonumber\\
&+&2\,\left( 2\,\pa_\rho B - \pa_\rho C \right) \,\left(
\frac{1}{\rho} + \pa_\rho C \right)  -
  \frac{\pa_\rho^2 A }{A} - 2\,\pa_\rho^2 C =0 \ .
\end{eqnarray}
Assuming that the interior equations are satisfied,  the Bianchi
identities ${\G_\alpha^\beta}_{;\beta} =0$, imply \cite{wis1} the
following relations between the constraint equations
\bea
\label{C-R}
&& \pa_\zeta {\cal U} +\pa_\xi{\cal V} = 0 \ , \nonumber\\
&& \pa_\zeta {\cal V} -\pa_\xi{\cal U} = 0 \ , \ena
where $\zeta=\log \rho.$, and we define the rescaled constraints ${\cal U} = \rho
\sqrt{-g} \left(\G_\rho^\rho-\G_\xi^\xi \right)/2 \ , {\cal V}=\rho^2
\sqrt{-g} \G_\rho^\xi$ with $g \equiv det\  g_{\alpha\beta}$.

A nice feature follows\cite{wis1}.  The constraints ${\cal U}$ and ${\cal V}$
satisfy the Cauchy-Riemann (\ref{C-R}) relations and hence each one of
them is a solution of the Laplace equation. Hence, if one of the
constraints is satisfied at all boundaries and the other at a single
point along some boundary these constraints must be satisfied
everywhere inside the domain. This fact will be referred hereafter as
the ``constraint rule''. In our implementation, following the choice in
\cite{wis1} we imposed ${\cal V}=0$ along all boundaries and ${\cal
  U}=0$ in the asymptotic region. It is important to check and confirm
that the constraint ${\cal U}$ and ${\cal V}$ are satisfied everywhere
for our numerical solutions, as we describe in section
\ref{sec_tests}.
\subsection{Boundary conditions and constraints}
\label{sec_BC}
The interior elliptic equations (\ref{EqA}-\ref{EqC}) are subject
to boundary conditions. In this section we describe the boundary
conditions that define the problem completely. The integration
domain is defined by $\{(r,z): 0 \le z \le L, ~r^2+z^2 \ge
\rho_h^{~2} \}$, designated by the thick dashed line in Fig.
\ref{fig_map}. The boundary conditions are specified on the axis,
at the horizon, in the asymptotic region and at the reflecting and
periodic boundaries $z=L $ and $z=0$.

\subsubsection {The $z=0$ and $z=L$ boundaries.}
On the reflecting, $z=0$, and the periodic, $ z=L$, boundaries we
impose
\beq \pa_z \psi=0,~~ \psi =A,B,C .
\enq
While at the reflecting
boundary this condition is simply $\pa_\xi=0$, at the periodic
boundary its implementation is not direct, see section
\ref{mesh_sec}.


\subsubsection{The $r=0$ axis.}
\label{sec_axisbc}
Regularity of the metric on the axis  (absence of a conical
singularity) requires
\beq
\label{Baxis}
B=C.
\enq
We use this equation as a Dirichlet condition for $B$. Equation
(\ref{EqB}) is not solved at the axis but it is only monitored there.
For $A$ and $C$ the boundary conditions are automatic -- on axis the
(interior) equations for these functions become first order in
derivatives normal to the boundary and have precisely the form of a
b.c. Namely these equations are already incorporate b.c. and these do
not need to be additionally specified.  We term this an ``automatic boundary condition''.  This occurs
because of our particular choice of the angular coordinate: we use
$\xi$ instead of $\chi$.  The axial symmetry of the problem dictates
the $\pa_r =0$ condition for the metric functions, which translates to
$\pa_\chi=0$ in spherical coordinates and $ \sqrt{1-\xi^2} \pa_\xi=0$
in our coordinates. But on axis $\xi=1$ and hence this condition need
not be imposed in our coordinates.  While the coordinate singularity
at the axis is quadratic, $\sim \sin(\chi)^{-2}$ when using $\chi$, it
becomes linear $ \sim (1-\xi)^{-1}$ when using $\xi$.  With this
advantage there is, however, a drawback: the metric functions are not
differentiable at $\xi=1$. This requires a modification of the
numerical scheme there, replacing the second order normal derivative
of the interior equations by a first order one due to the
considerations above as described in subsection (\ref{subsec_scheme}).

\subsubsection{The horizon}
\label{horbc}
The horizon in our construction is located at $\rho_h=1$. For
static solutions various notions of the horizon coincide -- the
event horizon (globally marginally trapped ), the apparent horizon
(the outermost boundary of locally trapped surfaces) and the
Killing horizon are all the same. The latter characterizes the
horizon as a surface where
\beq
A\equiv0.
\label{hor_A}
\enq
This implies that along the horizon
\beq
\pa_\xi A= \pa_\xi^2 A =0.
\enq
Even though the horizon normally is not singular (in curvatures) our
equations do become singular there as the function $A$ vanishes.  Now
we describe how the physical regularity of the equations at the
horizon gives boundary conditions for our functions\footnote{We assume
  hereafter that $\pa_\rho A|_{\rho_h} \neq 0$, i.e. the horizon is
  not degenerate.}.
Expanding  Eqs. (\ref{EqB},\ref{EqC}) at the horizon we obtain the
condition
\beq
\label{hor_C}
\pa_\rho C=-1 \ .
\enq
We still need a condition for $B$. We obtain this condition from
the zeroth law of the black-hole mechanics (or thermodynamics),
namely that for static solutions the surface gravity must be
constant along the horizon  (see for example \cite{wald}). The
surface gravity along the horizon reads
\beq \label{kappa} \kappa=e^{-B} \pa_\rho A, \enq
and the derivative of $\kappa$ along the horizon vanishes
\beq
\label{dkappa}
\pa_\xi \kappa \sim \pa_\xi B - \frac{\pa_{\rho\xi} A}{\pa_\rho A} =0,
\ \rm{at} \ \rho=\rho_h.
\enq
The upshot is that the boundary condition for $B$ can be obtained in
one of the forms: either
\beq
\label{hor_B2a}
B=C_{\xi=1}+\log{\left(\pa_\rho A \over \pa_\rho
    A_{\xi=1}\right)}\mid_{\rho_h},
\enq
 from (\ref{kappa}) and (\ref{Baxis}), or by integrating
 (\ref{dkappa}) outwards from the axis along the
horizon.  In our implementation we used the former form. However
we have checked that a corresponding solution obtained by using
the other option differs only slightly from our original one.

Note that the condition (\ref{dkappa}) implies that  eqn.
(\ref{GRchi}) (or ${\cal
  V}=0$) is guaranteed along the horizon, and vice versa.

We can get a different condition for $B$ as well. Examining
 eqn. (\ref{GRR-Gchichi}) (${\cal U}=0$) one obtains the condition
\beq
\label{hor_B1}
\pa_\rho B=-1 \ , \enq which is necessary to ensure regularity of that
equation along the horizon.

Altogether we now have too many conditions at the horizon: four
boundary condition
(\ref{hor_A},\ref{hor_C},\ref{hor_B2a},\ref{hor_B1}) for the three
metric functions. However, as explained in \cite{wis1} it is {\it
  unnecessary} to impose both constraints (${\cal U}=0$ and ${\cal
  V}=0$) at the same boundary, and actually it is {\it necessary not}
to impose both in order to protect the problem from being over
constrained.

Out of (\ref{hor_B2a},\ref{hor_B1}) we choose to impose
(\ref{hor_B2a}). The condition (\ref{hor_B1}) is satisfied for these
solutions. The error becomes smaller with grid refinement and reaches
$2\%$ for our finest grid. For the sake of completeness we also
obtained solutions using (\ref{hor_B1}) instead of (\ref{hor_B2a}).
However, these solutions do not have a manifestly constant surface
gravity.  The variation of $\kappa$ along the horizon is small for
small $x$, but can reach as much as $15 \%$ for larger $x$ values. The
overall difference between the two solutions is maximal near the
horizon, being of the same $15 \%$ magnitude. This difference fades
off asymptotically and the constraints are still satisfied (with the
same accuracy). We conclude that, in principle, it is possible to use
the condition (\ref{hor_B1}) to generate solutions, though the
numerics should be refined further to reach an acceptable accuracy.
%
\subsubsection{The asymptotic  boundary.}
Performing a Kaluza-Klein reduction one observes that the $z$-dependence
of all fields is carried by massive KK-modes and hence fades off
exponentially for large $r$. Thus, in the asymptotic region we can
rewrite Eqs. (\ref{EqA}-\ref{EqC}) retaining only the $r$-dependence.
Defining for convenience $\zeta\equiv log(r)$, we get
\bea
\label{asEqs}
&&A''+A'+ 2 A' C' = 0, \nonumber\\
&&B''-B' -2 C'-(C')^2 -{2 A'\over A} (C'+1) -1 +e^{2 B- 2 C}=0,\nonumber\\
&&C''+3 C' +2 (C')^2 +{ A'\over A} (C'+1) +1 -e^{2 B- 2 C}=0. \
\ena
Here the derivatives are calculated with  respect to $\zeta$.

Asymptotic flatness at $r \rightarrow \infty$, requires $A-1=B=C=0.$
Linearizing the above equations we can solve them analytically (see
appendix \ref{appendix_asympt}) with these boundary conditions
obtaining
\bea
\label{asflat}
A &=& 1-{a\over r} +{\cal  O}({\log^2 r \over r^2})\ , \\
B &=&{b \over r}+ {\cal  O}({1\over r^2})\ , \\
C &=&{c \log(r) \over r} + {\cal O}({1\over r})\ , \ena
where we also included the order of the corrections. Note the
logarithmic term in C. This log-behavior is specific to 5d and
indicates a very slow asymptotic fall-off.  At leading order the
coefficients in (\ref{asflat}) are related by \cite{numericI}
\beq
\label{coefficients}
a - 2 b +c =0 .
\enq

In the numerical solution one can impose the simplest Dirichlet
conditions: $A-1=B=C=0$ at the asymptotic boundary. However, since in
our numerical implementation the ``infinity'' boundary is located at a
finite $r$ this option appears to be too crude. One can improve that
by going to the next order in the expansion (\ref{asflat}) and using
(\ref{coefficients}) to get the refined conditions
\bea
\label{asflat1_a}
 && {d \over dr }(A r)=1  \ , \\
\label{asflat1_b}
 &&{d (B r )\over dr}=0 \ , \\
\label{asflat1_c}
&& C=(-1+A+2 B)\log(r) \ .
\ena
Here we have rewritten the conditions in a form convenient for
a numerical implementation.

Unfortunately, we discovered that these linear conditions do not
lead to a convergent scheme.  To understand this, observe  that
the function with the slowest decay is $C$. In 5d, to resolve the
difference between the first two terms in its asymptotic expansion
with just $10\%$ accuracy one has to go to $r \sim exp(10)\rho_h$
-- the log strikes hard!  When the maximal $r$ is not extremely
large (which is the case here for practical reasons) the
non-linear corrections appear to be important for stabilizing the
scheme \cite{wis1}.

Recall the ``constraint rule'' which will help us to derive a more
subtle b.c. for C.  In accordance with it we choose to enforce ${\cal
  V}=0$ along all boundaries.  The rationale behind it is that this
constraint is satisfied trivially at the axis and at the
reflecting boundaries, asymptotically it decays exponentially
fast, and only at the horizon this constrain is not trivial and
yields (\ref{hor_B2a}). The second constraint must not be imposed
at the horizon. At the axis it vanishes. Along the reflecting,
$z=0$, and the periodic, $z=L$, boundaries this constraint does
not carry any new information as it is just a linear combination
of the interior equations.  Hence, we are left with the asymptotic
boundary.  This boundary can be potentially dangerous since on one
hand, $ \left(\G_\rho^\rho-\G_\xi^\xi \right)$ decays here and on
the other hand the measure $\rho \sqrt{-g}$ blows up. This
competitive behavior can result in an unpredictable ${\cal U}$.
Thus, to guarantee that the constraint is satisfied, the natural
and unique place to impose ${\cal U}$ is the infinity.

The upshot is that instead of the linear condition
(\ref{asflat1_c}) we compute $C$ at the asymptotic boundary using
the constraint equation ${\cal U}=0.$ By doing so we stabilize our
algorithm and satisfy the ``constraint rule".  Note that at the
leading order the vanishing of ${\cal U}$ is consistent with the
linear condition (\ref{asflat1_c}).
%
%
%
\section{Thermodynamical and Geometrical Variables}
\label{sec_theory}
In this section we briefly summarize the results from our previous
paper \cite{numericI} with a particular focus on the 5d case.
Hereafter we work in units such that $G_4=1$.

\sbsection{At infinity}

As stated in \cite{numericI} (see also \cite{HO2}) there are two
energy-momentum charges that asymptotically characterize our
configuration. These are the total mass, $m$ (or the dimensionless
mass $\mu:=m/\hL$), and what we called the tension. The latter is what
an observer at infinity interprets as the tension of an imaginary
string stretched along the compact circle.  These charges can be
calculated in terms of the numerical asymptotics \cite{numericI}
\beq
  \hL \,  \left[ \begin{array}{c} \mu \\ \tau \\  \end{array} \right] =
  {1 \over 2\, } \left[ \begin{array}{cc}
  2 & -1 \\
  1   & -2 \\
   \end{array} \right] \,
 \left[ \begin{array}{c} a \\ b \\ \end{array} \right] .
 \label{asymp_to_charges}
 \enq
 The asymptotic mass can be calculated also in a direct way, using the
 free energy or the Hawking-Horowitz prescription\footnote{ The
   Hawking-Horowitz mass coincides with the ADM mass when both are
   applicable} \cite{HawkingHorowitz} which gives the same result.  In
 the linear regime, using the relation (\ref{coefficients}) in the
 above formulae one can express the physical charges in terms of any
 two of the numerical asymptotics.

\sbsection{At the horizon}

The characteristic quantities at the horizon are the surface
gravity (the temperature) and the area (the entropy) of the
horizon.  We have already defined the surface gravity in
(\ref{kappa}), and the temperature is proportional to it
$T=\kappa/2\pi$. The entropy is related to the surface area by the
famous Bekenstein-Hawking formula $S_{BH}=(A /4 G_N)$.  In our
coordinates the surface area reads
\beq
\label{3area}
A_3 = 4 \pi \rho_h^3 \int_{-1}^1{e^{B+2 C}}\sqrt{1-\xi^2} d\xi  = 4 \pi (2L)^3  x^3 \int_{-1}^1{e^{B+2
    C}} \sqrt{1-\xi^2} d\xi \ .
\enq
Out of these thermodynamical variables a single dimensionless
quantity can be formed
\beq
A^{(\kappa)} := A_3 \, \kappa^3 .
\label{Akappa3}
\enq
In addition to the 3-area it is useful to define a pair of 2-areas of
horizon sections.  The equatorial 2-area section is given by
\beq \label{Aeq} A_{\parallel}=4 \pi  \rho_h^2 e^{2C}. \enq
The 2-area of the section of the horizon along the axis is just
\beq \label{Aperp} A_{\perp}=2 \pi \rho_h^2 \int_{-1}^1 { e^{B+C}} d\xi.
\enq
With these 2-areas we define the eccentricity\footnote{This definition
  differs from the standard definition of an ellipse's eccentricity.
  It is analogous to $a/b-1 = (1-e^2)^{-1/2} \simeq 1 + 0.5\, e^2$,
  where $e$ is the conventionally defined eccentricity.}  or the
``deformation'' of the horizon as
\beq \label{epsilon} \epsilon ={A_{\perp} \over A_{\parallel}}-1
. \enq
Finally we define ``the inter-polar distance'' which is the proper
distance between the ``north'' and the ``south'' poles of the black
hole calculated along the axis. This distance reads
\beq
L_{\mbox{poles}} = 2 \int_{\rho_h}^{L} dz\, e^{B} ,~~ {\rm
  at} \ r=0.
\label{polardistance}
\enq
%

\sbsection{Small black holes}

For small black holes ($x \ll 1$) the
tension should vanish, $\tau\simeq0$, according to Myers \cite{Myers}. In this case we have
 \beq
 \left[ \begin{array}{c}
  a \\ b \\ \end{array} \right] \simeq  \left[ \begin{array}{c}
  4/3\\ 2/3\\ \end{array} \right] m ,  \label{small_asymp}
\enq
In addition, in this limit we have
\beq \mu \simeq{3 \pi \over 8} x^2.  \enq
Small black holes are expected to resemble a 5d \Sch
\ black hole for which we have
 \beq
\label{5dSch}
\rm{5d ~\Sch :} ~~~~~~  A_3^{Sch}=16 \pi^2 \rho_h^3, ~~~  A_2^{Sch}=16
\pi \rho_h^2, ~~~ \kappa={1\over 2 \rho_h}. \enq
More generally, the dimensionless variables can be expanded in a
Taylor series as a function of $x$.  It can be shown
\cite{GorbonosKol} that this expansion for $A^{(\kappa)}$ takes the
form \beq A^{(\kappa)}= 2\, \pi^2\, ( 1-3\cdot \zeta(2)\, x^2 +
\dots)~,
\label{Akappa3expansion}
\enq
where $\zeta(2)=\pi^2/6$ is the Riemann zeta function.

The analogous expansion for the eccentricity  reads
\beq
\eps = {8\over 3} \zeta(4)\, x^4 + \dots \ .
\label{epsexpansion}
\enq
Thus the prediction is that $\eps$ is positive, i.e. the black hole
becomes prolate along the axis.  This agrees with the intuitive
expectation that the black hole should approach a string shape as it
grows.

Other dimensionless quantities are
 the 3-area
\beq
\label{3area1}
{\cal A}_3 := {A_3 \over \hL^3}= 2 \pi^2 x^3  +\dots ,
\enq
the surface gravity and the temperature
\beq
\label{kappa1}
\bar{\kappa} := \kappa \hL = x^{-1}+\dots ~~ {\rm and} ~~ \bar{T} =
2\pi x^{-1} +\dots\ .
\enq
and the polar distance
\beq
\ell := L_{\mbox{poles}}/\hL.
\label{polardistance1}
\enq

\sbsection{Smarr's formula}

The Smarr formula, also known as the integrated first law, is a
relation between the thermodynamical variables of the problem both at
the horizon and at infinity.  It can be obtained either from the
(differential) first law together with scale invariance, or by
computing the Gibbons-Hawking free energy and combining it with the
expression for the mass\footnote{See \cite{ScalarCharge} for the
  appearance of the scalar charge in the first law}.  We find \cite{numericI}
\beq
\label{freeenergy}
 A_3 = { 8 \pi \over \kappa } a L .
\enq
This formula relates the horizon characteristics $A_3$ and $\kappa$
with the asymptotic variable $a$, together with the dimensionful
parameter $\hL$. This formula is an important test for our numerical
solutions.

\sbsection{A phase transition}

One of the most important questions that we aim to answer is whether
there is a maximal (dimensionless) mass of the black hole phase as
anticipated in \cite{barak1}. This is analogous to the minimal mass
$\mu_{GL}$ of the uniform black string below which the string is
classically unstable. What happens to a black hole more massive than
the critical black hole is unknown and constitutes one of the puzzles
of this system. The appearance of a critical mass should be signaled
in the numerics by a very slow convergence and ultimately no
convergence.

Given the asymptotic mass (\ref{asymp_to_charges}) of the black hole we can
calculate the area of the corresponding black string of the same mass
\beq
\label{BSarea}
A_{BS} = 4\, \pi\, (2\, G_4\, m)^2 \hL .
\enq
or in a dimensionless form
\beq
\label{BSarea1}
{\cal A}_{BS} := {A_{BS} \over \hL^3} = 16 \pi \mu^2   .
\enq
While the existence of a maximal mass designates a perturbative
(tachyonic) instability, the solution with ${\cal A}_{BS}={\cal A}_3$ 
designates the point of
the first order transition between the black hole and the black
string phases. This transition can occur quantum mechanically, via
tunneling, or by thermal fluctuations.

\sbsection{Summary}

In our problem we define four thermodynamic measurable
quantities namely the horizon 3-area, the surface gravity, the
asymptotic mass, and the tension, as well as two geometric quantities
the eccentricity of the horizon, and the inter-polar distance. The
thermodynamic ones are related by the very non-trivial Smarr formula
(\ref{freeenergy}). The validity of this formula for our measurables
is one of the most important tests for the numerics (in addition to
usual numerical tests of convergence, constraint violation etc.) This
is because the Smarr formula relates horizon variables, with
asymptotic ones. Hence the degree of violation of this formula can
serve as an indication of the global accuracy of the numerical method.
Another important task is to check whether the measurables have the
small $x$ asymptotics as expected/derived theoretically.  We use
dimensionless measurables: in this form the relations between them
remain the same regardless of whether $\rho_h$ or $L$ are
varied\footnote{For a fixed numerical lattice spacing it is not the same to
  vary $\rho_h$ or $L$.}.

\section{Numerical Implementation}
\label{sec_numerics}
In this section we describe our numerical algorithm for solving
the system of partial non-linear elliptic equations (\ref{EqA}-\ref{EqC}).
We estimate the rate of convergence of the numerics and check that
the numerical errors are small and the constraints are satisfied.
All our simulations were written in Fortran. The typical run time on
a 2GHz Pentium4 PC took about 1-2 days. The output of the code was
analyzed and visualized using Matlab.
\subsection{The scheme}
\label{subsec_scheme}
The numerical technique that is often implemented to solve partial
elliptic equations is an iterative method, called 'Relaxation', see
e.g. \cite{ames,NumRec}. In this method the solutions are
iteratively corrected, starting from some initial guess, until a
desired accuracy is reached. For non-linear equations a modification
is needed. Often an iterative Newton procedure is combined with
relaxation to find  solutions of non-linear equations
\cite{ames,NumRec}.

\subsubsection{Numerical lattice and discretization}
\label{mesh_sec}
Near the horizon we employ polar coordinates $\rhoxi$.
Asymptotically, however, cylindrical coordinates are the natural
choice. In order to use both we choose to divide our integration domain into two parts: (i) 'The nearby region' near the
horizon is covered by polar coordinates. (ii) 'The asymptotic region'
is glued to the nearby domain from the outer, far side and is covered
by cylindrical coordinates.  The two patches overlap in order to
exchange information about the functions during the relaxation.

We discretize our equations on a lattice that covers the domain of
the integration. We employ finite difference approximation (FDA) in
which one replaces  derivative operators by their discrete
counterparts. The discrete operators are obtained by a formal Taylor
expansion of functions at the grid points.  We use an FDA which is
second order in the grid spacing.  For example, if $h$ is the stepsize
in, say, the $\rho$ direction, which is sampled by index $j$, then the first
and the second derivatives of a function $\psi$ at the lattice point
$(k,j)$ are written to second order as
\beq
\label{fda}
\pa_\rho\psi_{k,j}={ \psi_{k,j+1} - \psi_{k,j-1} \over 2 h } +{\cal O}
(h^3), \ \pa_\rho^2\psi_{k,j}={ \psi_{k,j+1} -2 \psi_{k,j}+
  \psi_{k,j-1} \over h^2} +{\cal O} (h^3) .
\enq
Analogous expressions can be found for all other derivatives.

Our second order FDA incorporates a 5 point computation molecule
in the interior.  Obviously, it would be nice to have the same
feature also at the boundaries. In fact, we retain this feature at
all boundaries but the axis through the introduction of false grid
points.  For the functions that have a Dirichlet boundary
condition we solve inside the domain using data at the boundaries.
This is the method for A and B on the horizon and for C
asymptotically.  To implement a Neumann condition we introduce
false grid points located one stepsize outside the real
boundaries. Since the normal derivative is given at the real
boundary we define the function at the false grid points using the
corresponding inner points. For example, at the horizon:
  $\pa_\rho C = (C_{k,j_h+1}-C_{k,j_h-1})/2 \Delta \rho =-1$, where
  $j_h=1$ is the location of the horizon $\rho_h$; at the false point
  $(k,j_h-1)$ we have $C_{k,j_h-1}=C_{k,j_h+1}+2 \Delta \rho$.
   Now we solve the equation on
the real boundary just as for an interior point, using data at the
false grid points.  This is the method for C on the horizon and
for A, B and C on the equator.  The mixed Neumann-Dirichlet
conditions for A and B, that are written in the form
(\ref{asflat1_a}-\ref{asflat1_b}), are imposed in the same
fashion. At the axis we have ``automatic conditions'' for A and C.
Since the functions are not differentiable here in the $\xi$
direction we use one-sided (first-order) $\xi$-derivatives.  Here
our FDA, becomes 4 point and not completely second order as
(\ref{fda}), see also the discussion in section \ref{sec_axisbc}.

The boundary $z=L$ is rather complicated in polar coordinates
while being very simple in cylindrical ones. This brings us to a
closer examination of the coordinate patches.

{\it The nearby  region: $\{\rho,\xi\}$ - patch.}
%
\begin{figure}[t!]
\centering
\noindent
\includegraphics[width=14cm]{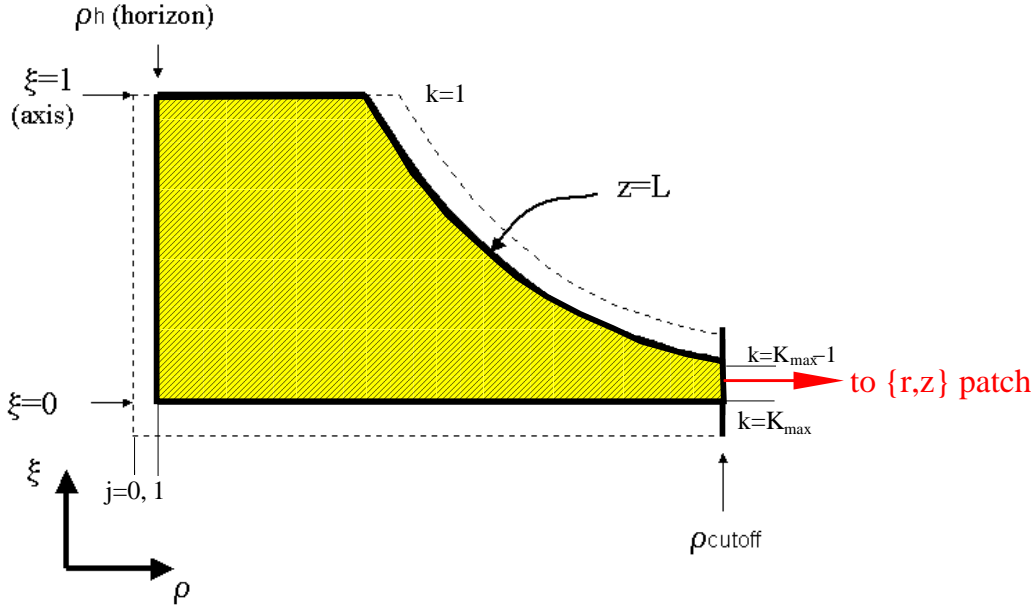}
\caption[]{'The nearby region' of the integration domain covered by
  the polar coordinates $\rho,\xi$. The thin dashed lines mark the
  location of the false grid points used for numerical implementation
  of the Neumann or mixed Neumann-Dirichlet boundary conditions.}
 \label{fig_domain}
\end{figure}
The lattice that covers this domain, see Fig.(\ref{fig_domain}), has
nodes at $k=1,...,K_{\rm max},K_{\rm max}+1$, in the $\xi$ direction,
and nodes at $j=0,1,...,J_{\rm in}(k)$ in the $\rho$ direction.  Here
$J_{\rm in}(k)$ is the coordinate of the last point that lies { \it
  within} the boundary $z=L$, which is represented by the curve
$\rho=L/\xi$.  The false grid points here are introduced by the
requirement that for each inner point there will be outer points that
allow the implementation of the regular 5 point second order FDA
scheme.  This implies that at each $k$ the outer points occupy $ J_{\rm
  in}(k)<j \leq J_{\rm out}(k)$ where $J_{\rm out}(k)=J_{\rm
  in}(k+1)$.  During the relaxation we sweep the lattice from $k=1$ to
$k=K_{\rm max}$ and from $j=0$ to $j=J_{\rm in}(k)$.
At a given ray $k$ we correct the outer points
at $k-1$ when we reach $j>J_{\rm in }(k-1)$ using the reflection b.c. To this end, for each outer point,
${\cal O}$, we find the corresponding inner point, ${\cal I}$, which
is its mirror reflection with respect to the boundary ($(r,z) \to (r,2\, L-z)$).
The functions at the inner point are obtained by a two dimensional interpolation from the
surrounding points and then the corresponding outer point is updated.
The important practical note is that the inner mirror points should be
calculated only once prior to relaxation.

We choose fixed lattice spacings both in $\rho$ and $\xi$ directions.
In the $\rho$ direction the grid is truncated at a finite $\rho_{\rm
  cutoff}$. For a specific $\xi-$stepsize, $\Delta\xi$, the maximal
$\rho$ is
\beq
\label{rhocutoff}
\rho_{ \rm cutoff}={L \over \Delta\xi }.
\enq
The $\Delta\xi$ step is chosen such that $\rho_{\rm cutoff} \gg L$
(usually we took $\rho_{\rm cutoff} \sim 10 L$.)  Note that there are
only two grid points on this far boundary in the $\chi$ direction for
$\rho_{\mbox{cutoff}}$: one is at $K_{\rm max}-1$ and the other is at
$K_{\rm max}$, see Fig. \ref{fig_domain}.  Here the second patch of
the integration begins.

\begin{figure}[t!]
\centering
\noindent
\includegraphics[width=10cm]{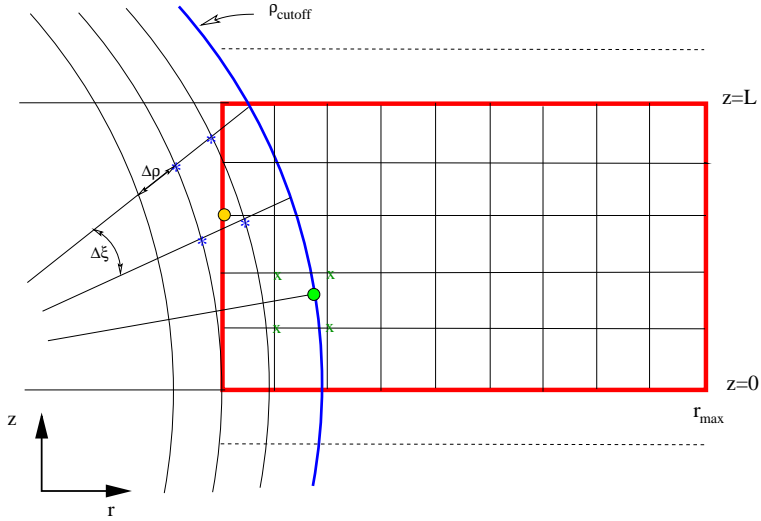}
\caption[]{The asymptotic region is glued to the 'nearby patch'. The
  two patches overlap in order to exchange information about the
  functions during relaxation.}
 \label{fig_rzpatch}
\end{figure}

{\it The asymptotic region: $\{r,z\}$ - patch.}
This patch begins at $r=r_{ \rm min} <\rho_{\rm cutoff} $ and extends
up to $r=r_{\rm max}$.  Note that there is a 'buffer zone' where both
patches overlap, see Fig.\ref{fig_rzpatch}.  The variation of the
functions in this portion of the integration domain is expected to be
small provided that $\rho_{\rm cutoff} \gg L$.  Thus, the lattice
covering this portion does not need to be very dense.  The grid has a
simple rectangular geometry with uniform grid spacings. There are two
false boundaries at $z=0,z=L$. At the near boundary, $r_{\rm min}$, all
functions have Dirichlet b.c., the values being received from the
'nearby patch'. At the far boundary, $r_{\rm max}$,we implement the
mixed Dirichlet-Neumann conditions (\ref{asflat1_a}-\ref{asflat1_b})
for A and B and evaluate C from the constraint ${\cal U} =0$.  Since
for practical reasons we were obligated to take finite, not too large
$r_{\rm max}$ (we usually chose $r_{\rm max} \sim {\cal O}(1000)\rho_h$)
the use of the non-linear condition for C is essential.

The equations are relaxed on both patches one after the other. First
we sweep the lattice of the 'nearby region' and then the lattice of
the asymptotic region. Then the sweeps are iterated.  The patches
communicate. When sweeping the near patch we use the information from
the far patch to supply boundary conditions at $\rho_{\rm cutoff}$.
For example, see Fig. \ref{fig_rzpatch}, the green point at $\rho_{\rm
  cutoff}$ can be obtained by e.g. a bi-linear interpolation from the
points marked by crosses. While sweeping the far patch the boundary
condition along the stitch at $r_{ \rm min}$ come from the near patch.
For example, the yellow point at $r_{\rm min}$ is obtained from the
points marked by stars by a bi-linear interpolation.

To relax the equations we used a scheme which incorporates
Newton iteration. To this end, at each grid point $(k,j)$, any function
$\psi = A,B,C$ is updated according to
\beq
\label{gausssiedel}
\psi^{\rm new}(k,j)=\psi^{\rm old}(k,j) - \omega{{\cal E}_\psi(k,j) \over
  \pa{\cal E}_\psi / \pa\psi(k,j)}\mid_{\rm old} ,
\enq
where ${\cal E}_\psi(k,j)$ is the FDA equation of motion for this
$\psi$, and $\omega$ is a numerical factor.

The basic Gauss-Seidel algorithm uses $\omega=1$ and it leads to a
very slow convergence\footnote{In fact, in our case this method is so
  slow that we could not infer that it converges at all.}. One way to
speed up the performance is to use Successive Over Relaxation (SOR).
The name originates from the fact that unlike the Gauss-Seidel scheme
where the functions are corrected by exactly what they should be from
the equations, in the SOR algorithm the correction is larger. The
over-relaxation is managed by the relaxation parameter $1<\omega<2$.
Sometimes, for non-linear equations it is better to use
under-relaxation. In this case $0<\omega<1$ and the functions are
under-corrected. The case $\omega>1$ ($\omega<1$) can be imagined as a
sort of acceleration (friction). We implemented the SOR algorithm for
our problem and found that the convergence rate is still
unsatisfactory for dense grids.

\subsubsection{Multigrid technique}
\label{muli_sec}

The algorithm that we found to work well is what we loosely term here
a Multigrid algorithm, see e.g. \cite{NumRec}.  In the current
simplified version of this method we solve the equations on several
successive grids with doubled density.  The basic idea of the
Multigrid technique is simple -- relax perturbations of different
wavelengths on suitable lattices.  Clearly, relaxation of a
long-wave perturbation on a very fine grid will require many iterations,
while if we first relax the perturbation on a course grid and use the
solution as an input for the fine grid there is a chance to converge
faster.  Another advantage of using the Multigrid method is that there
is a natural measure of accuracy.  We can compare the solutions on
different grids and see whether and how the differences decrease,
indicating convergence and scaling of the truncation error. We used
$4$ successive grids with halved stepsizes.  Our implementation of the
multigrid method was very simple. We just improved the solution going
in one direction, namely passing from a coarse grid to a finer one.
The original multigrid technique \cite{NumRec} incorporates motion in
both directions allowing to relax newly excited modes on suitable
grids.  This two-directional method is expected to be much more
effective and fast (but difficult to program.)

The Multigrid technique was implemented only on the 'nearby patch'.
The asymptotic patch was chosen to have fixed grid spacings. This is
because the variation of the fields is relatively small at large $r$
and there is no need for very dense grids.  Note also that by
decreasing the grid spacings in the nearby patch $\rho_{\rm cutoff}$
scales according to (\ref{rhocutoff}). Thus for each more dense grid
$\rho_{\rm cutoff}$ is doubled. We, however, chose to keep the
truncation radius constant, defined by the coarsest grid.  In this
case when the grid spacings were halved the number of
the points on the boundary at $\rho=\rho_{\rm cutoff}$ was doubled.

In most of our simulations typical grid spacings in the asymptotic
patch were $\Delta z \simeq 0.25 - 1.0$, and $\Delta r \simeq
0.075- 0.25$. The typical grid spacings in the near patch were
$\Delta \rho \simeq 0.1- 0.25$ and $\Delta \xi \simeq 0.08 -
0.12$ for the {\it coarsest} grid. One can estimate the size of the
lattice taking typical $L\sim 10\rho_h, ~\rho_{ \rm cutoff },r_{\rm
  min} \sim 10 L ~~ {\rm and}~~ r_{\rm max} \sim 10 \rho_{ \rm cutoff
  } \sim 100 L \sim 1000\rho_h$. In this case, the size, $N_{\xi}
\times N_{\rho}$, of the coarsest polar grid is $10 \times 1000$,
while the finest grid is $160 \times 16000$.  The size, $N_{z} \times
N_{r}$, of the asymptotic grid would have been $10\times 1000$, if the
grid spacing in the $r$ direction were uniform.

\subsubsection{Extracting measurables}
Once we have a solution the horizon variables such as $\kappa, A_3
,A_2$ are calculated in a straightforward way from (\ref{kappa}),
(\ref{3area}) and (\ref{Aeq},\ref{Aperp}) respectively .  In order to
obtain the asymptotic mass and the tension (\ref{asymp_to_charges}) we
have to expand the metric functions asymptotically. We use fitting
functions of a suitable form to obtain those coefficients.  For
example to find $b$ we need a quadratic fit in $r^{-1}$ for $B$ in the
asymptotic region. To find $a$ a linear fit is usually sufficient for
a good result. For $C$ we used a fit of the form $c_1\log(r)/r +
c_2/r$. However, we were unable to find a reliable fitting for $C$. We
associate this with the slow logarithmic decay.
\subsubsection{Further developments}
\label{sec_tweaks}
There are always compromises in numerics between the computation time
and the accuracy of the calculation. This has to do with the grid density.
Large grids mean small stepsizes and hence better accuracy provided
that the FDA is stable, i.e. converges as a stepsize decreases. On the
other hand, even if there were unlimited memory resources to store
large arrays, such dense grids would result in extended CPU time that
would be needed to sweep such large lattices. We tried different
tricks to find a reliable compromise for this 'CPU-time vs. accuracy'
issue.

$\bullet$ {\it A non-uniform asymptotic grid.} Originally we
implemented the asymptotic boundary conditions at $r_{\rm max}$, that
is 30-50 $L$, or about 300-500$\rho_h$. However, we found that those
values of $r_{\rm max}$ are not large enough to determine the mass
with sufficient accuracy, especially for large values of $x$.  Since
in 5d the asymptotic fall-off is slow, (\ref{asflat}), implementing
the asymptotic boundary conditions at these $r_{\rm max}$ values is
not accurate enough. 
Naturally, one would like to extend the grid to
the larger $r_{ \rm max}$. One way to do so is to add more points to
the grid. However, in order to reach $r_{ \rm max}$ larger by a
certain factor than the original $r_{ \rm max}$ the number of grid
points must increase by the same factor. The same is true for the
increase in the CPU time. Since we need $r_{ \rm max}$ of order
$\exp(10)\rho_h$, this makes the mere increase in number of points
absolutely impractical.

We used another technique -- a non-uniform grid spacings in the
r-direction.  The stepsizes were scaled in the following fashion
\beq
\label{nonunstep}
\Delta r_{i+1} =(1+\epsilon) \Delta r_i \ , \ i=1,2,\ldots,N_r
\enq
Here $\epsilon$ is a small number that we usually took as $0.01 -
0.03$.  In this case the coordinates of the mesh points in the $r$
direction form a geometric progression and it is possible to reach
quite large values of $r_{\rm max}$ with a relatively small number of
the mesh points.  The discretized equations are now modified and the
truncation error at a grid point $i$ scales as ${\cal O}( \eps
\Delta r_i)+{\cal O}( \Delta r_i^2)$ rather than just ${\cal O}( \Delta r_i^2)$, which is the case
for a uniform grid with the spacing $\Delta r_i$. Provided $\epsilon$
is small this modification is not that different from a uniform FDA
and does not cause problems.  In the $z$-directions stepsizes were left
uniform and hence the corresponding FDA derivative operators remained
unchanged.

Keeping $\epsilon$ in the above range, in order to reach large enough
$r$ (of order $\exp(10)\rho_h$)the grid turns out to be large enough to
slow down the convergence notably. In practice, we could reach only
$r_{\rm max} \simeq 500-800 \rho_h$. The log strikes again.

$\bullet$ {\it Additional relaxation near the horizon.}  This is
needed in order to increase the accuracy of the calculation of
$\kappa$ and the area of the horizon and its sections
(\ref{3area}-\ref{Aperp}).  Especially, the eccentricity
(\ref{epsilon}) is sensitive to the accuracy of the area measurement.
This relaxation operates over a finite portion of the mesh in the
vicinity of the horizon. The boundary conditions along the horizon,
axis and the equator are the same as before but along the outer
boundary one uses just the Dirichlet boundary conditions that come
from the main relaxation.

Over this region the metric functions were relaxed on two additional
finer grids.  One could suspect that the relaxation over a finite
region would produce a mismatch along the stitch, that can be imagined
as a kink or a 'ripple' in the metric fields.  However, we have
checked that this is not the case, but rather the behavior of the
functions was smooth. The maximal change in the functions relative to
the previous grids occurred over a few mesh points near the horizon.

In addition, in a couple of runs we relaxed the equations over the
{\it entire} integration domain on an additional 5th grid. When the
area and the surface gravity, obtained in this case, were compared to
the ones obtained in the relaxation over only a small region near the
horizon, we found that the difference in both results is less then 0.1
\%. The gain in computation time was however dramatic -- hours vs.
days.

$\bullet$ {\it The over-(under-) relaxation parameter $\omega$.}
There is no universal algorithm to find the optimal $\omega$ that
speeds up the convergence.  Except for few very simple elliptic
equations with simple boundaries there is no analytical prescription
to pick such an optimal $\omega$. Often empirical estimates are the
only way to find it. In our case the estimation of an optimal $\omega$
is even harder, since we need an omega for each of the three functions
for both patches.

While we cannot be confident that the $\omega$-s we have used in our
computations are the optimal ones, we can estimate the range of $\omega$ outside which the code slowed down or diverged.  The
choice $\omega_A=1-1.2, \ \omega_C=1-1.1, \ \omega_B=.5-1.$ in the
nearby region, and, $\omega_A=\omega_C=1-1.2, \ \omega_B=.02-0.1$ in
the cylindrical patch usually gave reasonable convergence rate. Some
of these values depend on $x$. The most influential $\omega$ is $\omega_B$
in the asymptotic region: a slight deviation of its value from the
narrow range ruins the convergence.
%
\subsection{Testing the numerics}
\label{sec_tests}
As usual in numerics one has to convince oneself that a particular
numerical method produces trustable results.  Before discussing our
findings, we present in this section the evidence that our method
performs well. We show that the numerical errors decrease sufficiently
fast indicating a global convergence of the scheme and that the
residual errors of the equations are small. In addition, in GR one has
to ensure that the constraint equations are satisfied.  We show that
they are satisfied to a good extent.  However, this ceases to be
the case when $x$ is 'too large'. For $x$ above a certain value $x_1\simeq
0.20$, the convergence is slowed down, the constraints are violated
significantly and the errors are not small  for the results to
be reliable. Additional accuracy estimates come from the Smarr
formula, which our solutions  satisfy with very good
accuracy.

\sbsection{Numerical tests}

\begin{figure}[t!]
\centering
\noindent
\includegraphics[width=8cm]{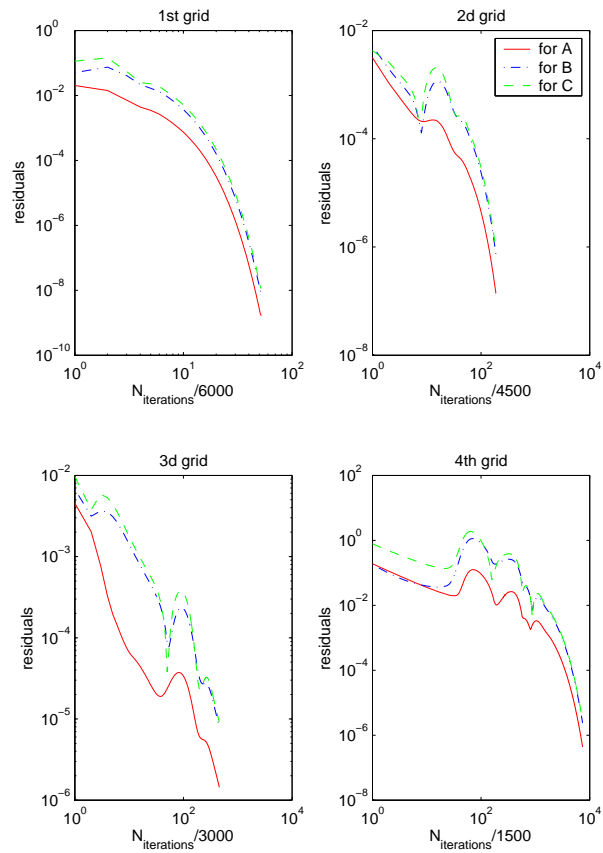}
\caption[]{A log-log plot of the normalized residuals,
  $Res \ \psi$ vs. number of iterations, for
$x=1/7$, implying  convergence.}
 \label{fig_residuals}
\end{figure}

We relax the equations on four grids with increasing density. For the
first and coarsest grid we give some initial guess while when moving
to a finer grid we start with the solution relaxed on the previous
grid.  The first thing that a good method must satisfy is independence
of the initial guess. To achieve faster convergence we regularly used
as the initial guess the uncompactified 5d \Sch \ solution restricted
to $|z|\leq L$. However, we checked that other initial guesses, such
as flat spacetime  glued to the horizon etc. relax to the
same final solution.  As an indicator of the accuracy during the
relaxation we use the accumulative residual error defined as
\beq
\label{residDef}
{\rm Res } \ \psi := {1\over 4^{n-1}}\sum_{k,j}{ \left| \triangle \psi_{k,j}
  -{Src\psi}_{k,j} \right|} \ , \rm{with} \ \psi = A,B,C \ ,
\enq
where $n$ is the grid number ($n=1$ is the coarsest) and the factor
 $4^{n-1}$ roughly compensates for the increase in the total number of
 grid points.
 The iterations on a particular grid were stopped when the residuals
 were reduced by a desired factor relative to some number, that we
 usually took as the initial residual calculated before relaxation on
 that grid.  In Fig.  \ref{fig_residuals} we depict the
 residuals on each grid.  Their behavior suggests convergence.
Note that the decrease of residuals is not monotonic all the way down.
This implies that there are modes that are continuously excited and
then decay.  For small values of $x$ these modes are harmless, but have an
imprint on the residuals' decay -- the oscillations. For larger $x$ values
these modes are not suppressed -- the oscillation increases their
amplitude and then they finally diverge, see Fig.{\ref{fig_divergence}.
\begin{figure}[b!]
\centering
\noindent
\includegraphics[width=7cm]{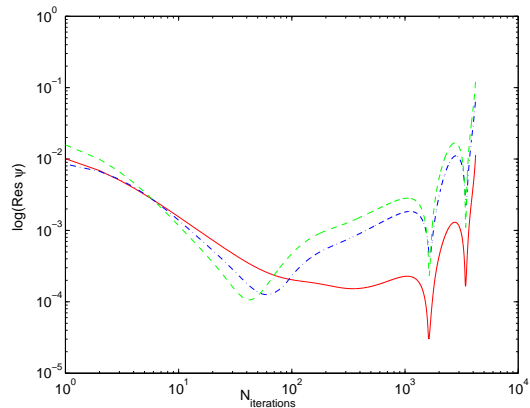}
\caption[]{A log-log plot of the  normalized residuals vs. number of
  iterations for $x>0.20$. After an initial convergence the solution
  diverges. }
 \label{fig_divergence}
\end{figure}

In addition to the convergence {\it per grid} we can use the benefits
of the multigrid technique and check the convergence when {\it moving
  between grids}. We examine how much the solution is corrected
when relaxed on different grids. To this end in Fig
\ref{fig_diffgrids} we depict the {\it differences} between the
solution on the $n$th grid and the solution obtained on the $n-1$,
coarser grid.  Since we used four grids there are three such
pairs. We observe that the solution is corrected less on finer
grids, as expected for a convergent method.  Most of the
corrections occur in the regions near the horizon and near the
axis.  In fact, this is the sort of behavior that allowed us to
perform further relaxation with confidence  even on finer grids in
the vicinity of the horizon as described in the previous section.
\begin{figure}[t!]
\centering
\noindent
\includegraphics[width=10cm]{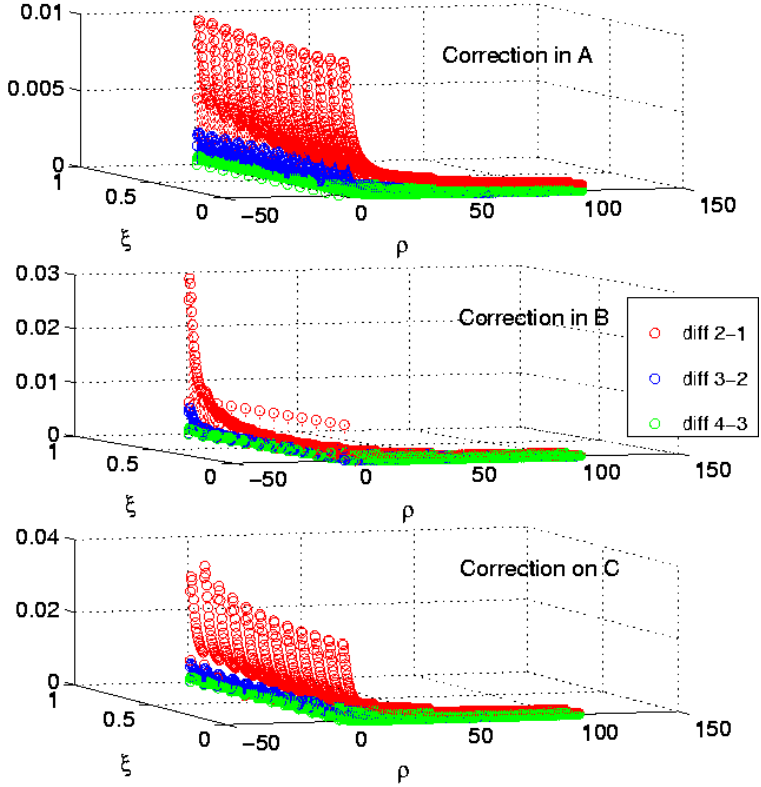}
\caption[]{ The corrections (or differences)  in the elliptic
  equations when relaxed on different grids for $x=0.1$. One can see
  that the finer the grid, the less the solution is corrected.}
 \label{fig_diffgrids}
\end{figure}

We  can also  estimate the {\it rate of convergence}. Assuming that the
solution converges to some $f^\star$ in the limit when the grid spacing
goes to zero, we can write  for the $n$th grid
\beq
\label{converg2}
f^\star=f_n+{\cal O }(h_n^p),
\enq
\begin{figure}[h!]
\centering
\noindent
\includegraphics[width=8cm]{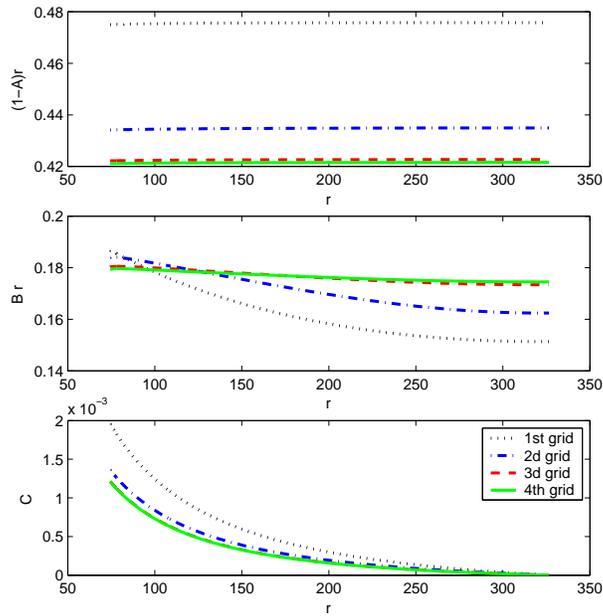}
\caption[]{A run with $x=0.12$. Values of the functions at the equator
  in the asymptotic region for 4 grids. There is clear convergence.  }
 \label{fig_converg2}
\end{figure}
where $h_n$ is the grid-spacing for this mesh.  The rate of
convergence is defined by the power $p$. If $p=1$ one speaks about
linear convergence, if $p=2$ there is a quadratic convergence and
so on. Taking differences between solutions on different grids and
considering the fraction of these differences the value of $p$ can
be estimated. Of course, the value of $p$ may vary at different
points of the numerical lattice. However, one can calculate the
{\it minimal } convergence rate by taking the minimum of those
$p$.  We find that in the asymptotic region the minimal
convergence rates are $p_A \sim 3, \ p_B \sim 2, \ p_C \sim 3 $
for the three metric functions. In the nearby region the
convergence rate is also found to be at least  quadratic for all
functions. In the above estimates of $p$ we used only the three
finest grids. The reason not to include the first grid is simple:
its prime role is to perform a rough adjustment of the initial
guess to the given boundary conditions. Hence, one expects that
the solution on this grid is only a very crude approximation to
the final solution.  To guide the reader we plot in Fig.
\ref{fig_converg2} the metric functions in the asymptotic region
at the equator for four grids. The nice convergence there can be
easily seen.  When this is the picture we infer that our method
converges nicely.
\begin{figure}[t!]
\centering
\noindent
\includegraphics[width=12cm]{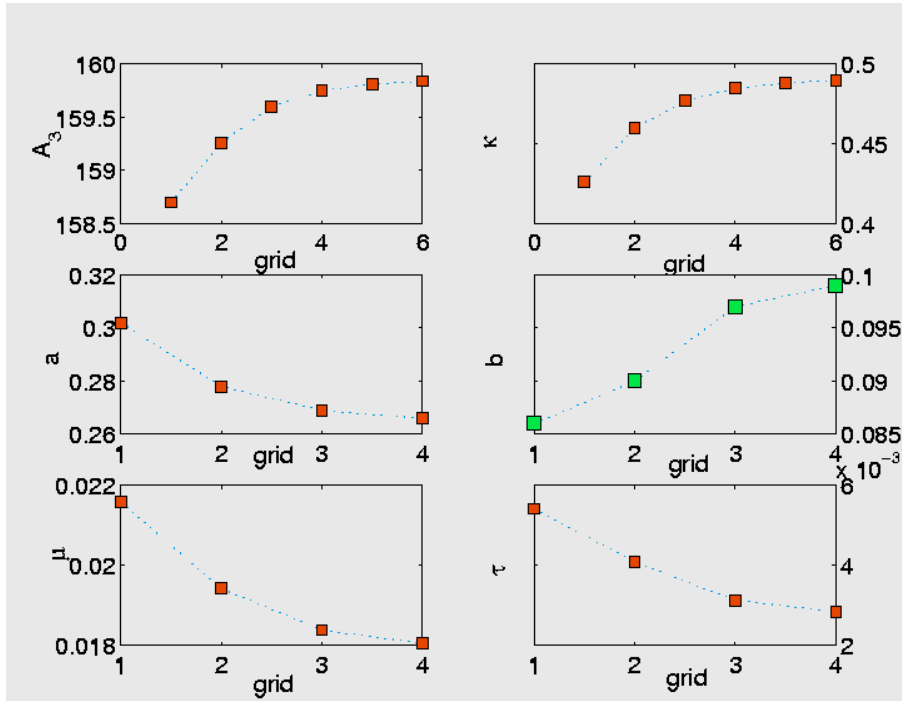}
\caption[]{ Four measurables and two asymptotics as a function of grid
  number
  for $x=1/12$. While the 3-area, $\kappa$ and $a$ converge nicely for
  all the $x$'s that we relaxed, $b$ does not. The absolute variation
  of the variables is small however. Note, however that the asymptotic
  charges, $\mu$ and $\tau$, converge as well.}
 \label{fig_converg1}
\end{figure}

Convergence of this kind occurred for intermediate values of $x$,
roughly for $0.08 \lesssim x \lesssim 0.15$.  Outside this range the
rate of convergence is still very good for all measurables but one. To
envisage our point it is more convenient to use the numerical
asymptotics $a$ and $b$, instead of our 'physical' measurables, the
asymptotic mass and tension\footnote{Note that $b$ does have a
  physical meaning -- it is the scalar (or dilatonic) charge.}.  The
typical behavior for our measurables as a function of the grid
spacing is depicted in Fig.  \ref{fig_converg1}. While $A_3,
\kappa$ and $a$ are observed to reach their asymptotic values, $b$
is special -- it does not seem to converge to a definite value.
Note that the asymptotic charges, $\mu$ and $\tau$ seems to settle
to definite values as well. Moreover, even though $b$ does not
behaves monotonically, for small $x$ its value is limited to be
within a narrow range, see Fig. \ref{fig_converg1}. We will see
below that the real problems of convergence begin when the
fluctuations of $b$ are not small anymore. We refer to these
fluctuations and the lack of accuracy in the measurements of $b$
as the ``the b-problem''.

To get an idea of the overall accuracy of the numerics in the final
solution, i.e.  after relaxation on all four grids, we plot in Fig.
\ref{fig_relerrors} the normalized error in our elliptic equations.
This error is defined at each mesh point $(k,j)$ as
\beq
\label{relErrors}
\delta \psi_{k,j} \equiv { {\triangle \psi}_{k,j} +{Src \ \psi}_{k,j}
  \over \left|{\pa_\rho^2\psi}_{k,j} +{{1\over \rho} \pa_\rho
      \psi}_{k,j}\right| +\left|{{1 \over
        \rho^2}\sqrtmu\pa_\xi\left(\sqrtmu \pa_\xi
        \psi\right)}_{k,j}\right| +\left|{Src \ \psi}_{k,j}\right| }
\enq
\begin{figure}
\centering
\noindent
\includegraphics[width=12cm]{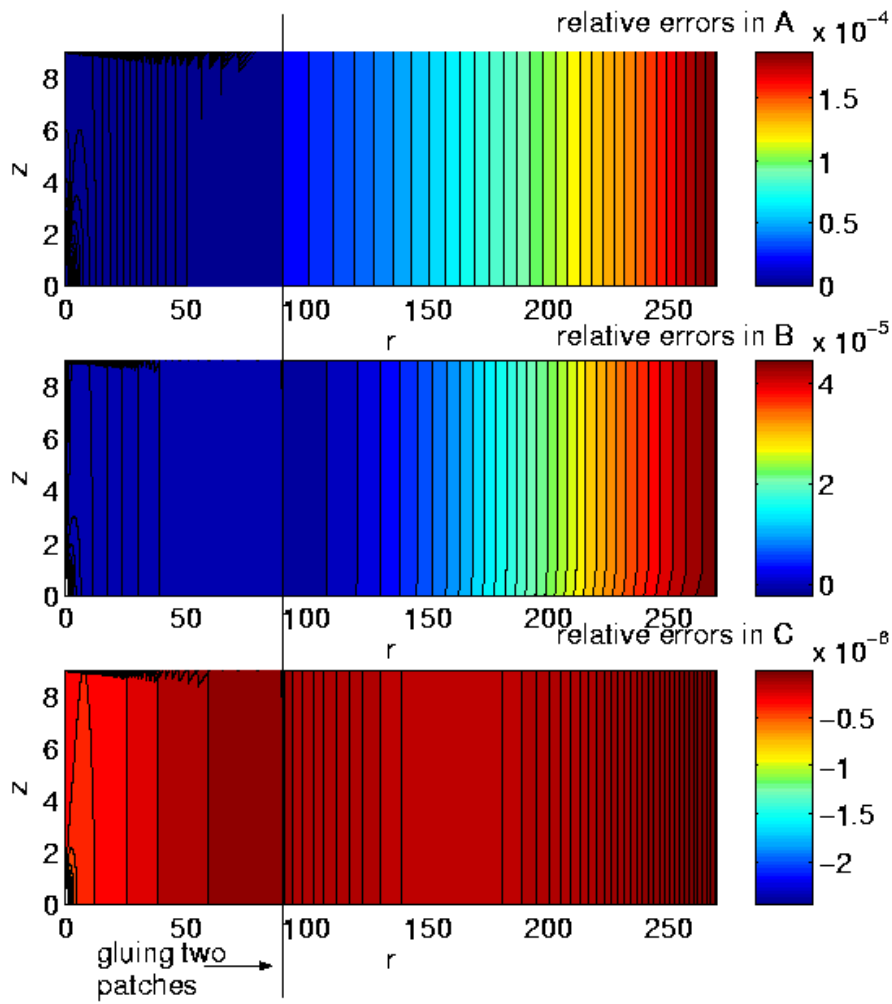}
\caption[]{ The normalized error in the elliptic
  equations, for $x=1/9$. This error defined in \ref{relErrors}. It is
  encouraging that the maximal error is less then 0.02 \%.  We plotted
  here errors in both patches.}
 \label{fig_relerrors}
\end{figure}
One observes that the relative errors are very small being less
then $0.02\%$. As $x$ approaches $0.20$ the errors grow to a level
of  few percent.

\sbsection{Constraint equations}

Additional insight into the accuracy of the method can be gained
by studying the behavior of the constraint equations. It is clear
that these must be satisfied for the actual solution of the
Einstein equations.  In Fig. \ref{fig_Con_abs} we plot the
absolute value of the constraints on both grid patches. The figure
shows that the constraints are not small in the far region of the
polar patch.  We believe that this loss of accuracy is due to the
geometric pathology of the polar coordinates in the asymptotic
region: the uniform grid cells in polar coordinates become very
thin and prolonged when viewed in Cartesian coordinates. This
causes a loss of accuracy, since asymptotically there are very few
mesh points in the polar patch. Hence, an attempt to compute
derivatives in the $z$-direction that are required from the
physical point of view, gives inaccurate result. When we passed
from grid to grid the constraints were satisfied better. When
evaluating the constraints in the Cartesian patch we did not
observe any pathologies. The constraints were small and decreased
fast in the asymptotic region.
\begin{figure}
\centering
\noindent
\includegraphics[width=12cm]{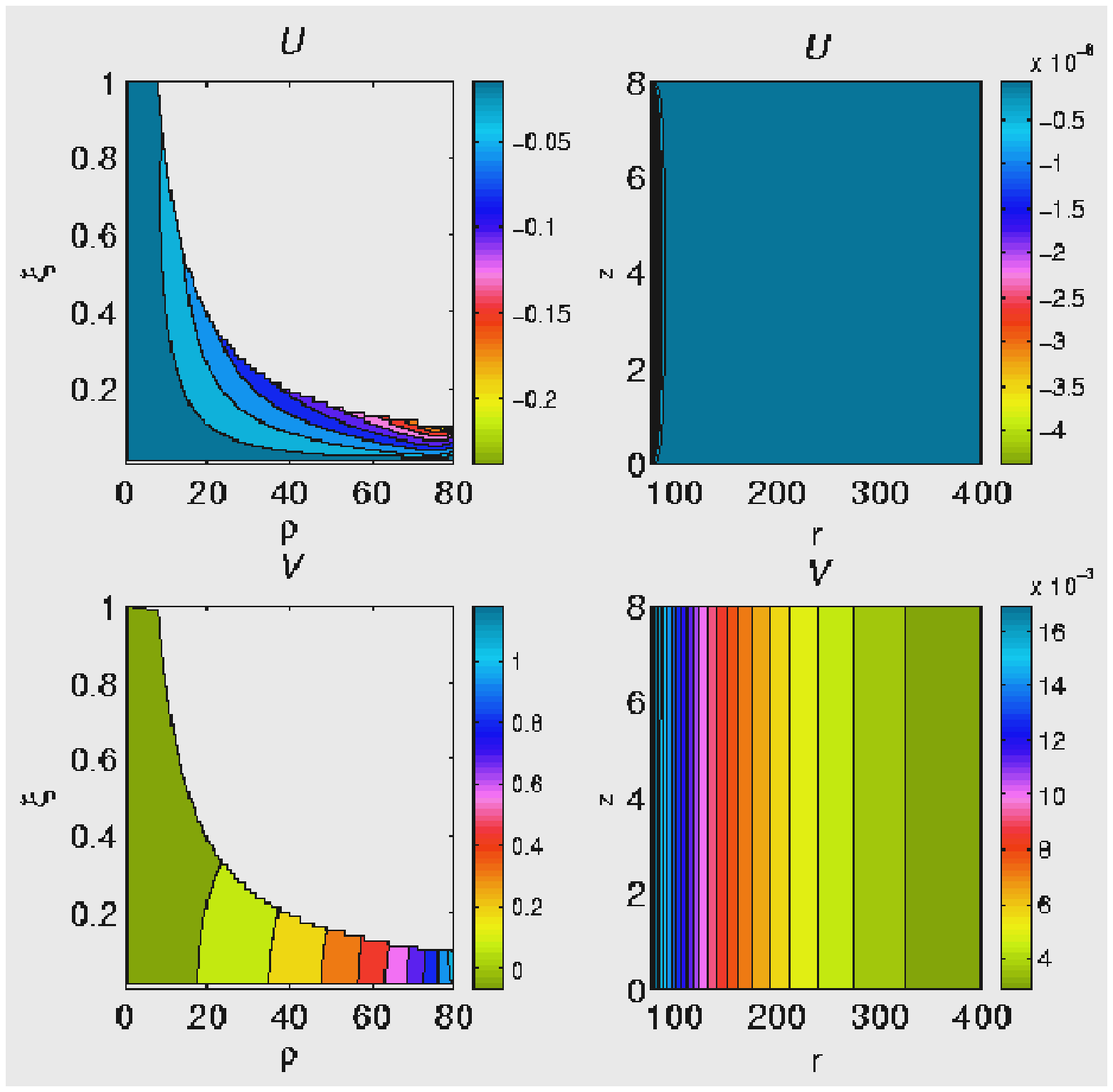}
\caption[]{Absolute value of the constraint
  equations on both grid patches. While there is a loss of accuracy
  near the far boundary  in the polar patch, both
  constraints are small in the Cartesian patch.}
 \label{fig_Con_abs}
\end{figure}

One gets a better insight for the constraints' accuracy from
examining the {\it relative} errors in them.  These  are defined
similarly to Eqn.  (\ref{relErrors}) and plotted in Fig.
\ref{fig_Con_rel}. One learns that the relative errors in the
polar patch are indeed very small, being less then one percent,
even though the absolute value of the constrains is not small.
Thus, both Figs. \ref{fig_Con_abs} and \ref{fig_Con_rel} are
complimentary and bring to light different aspects of the
constraints behavior.  The relative errors in the Cartesian patch
are difficult to calculate because of the small absolute values of
the constraints there. The absolute errors are small there and
vary slowly, see Fig.\ref{fig_Con_abs}. Hence an attempt to
evaluate the relative accuracy according to (\ref{relErrors})
fails, producing unpredictable results because of  roundoff
errors.
\begin{figure}[h!]
\centering
\noindent
\includegraphics[width=12cm]{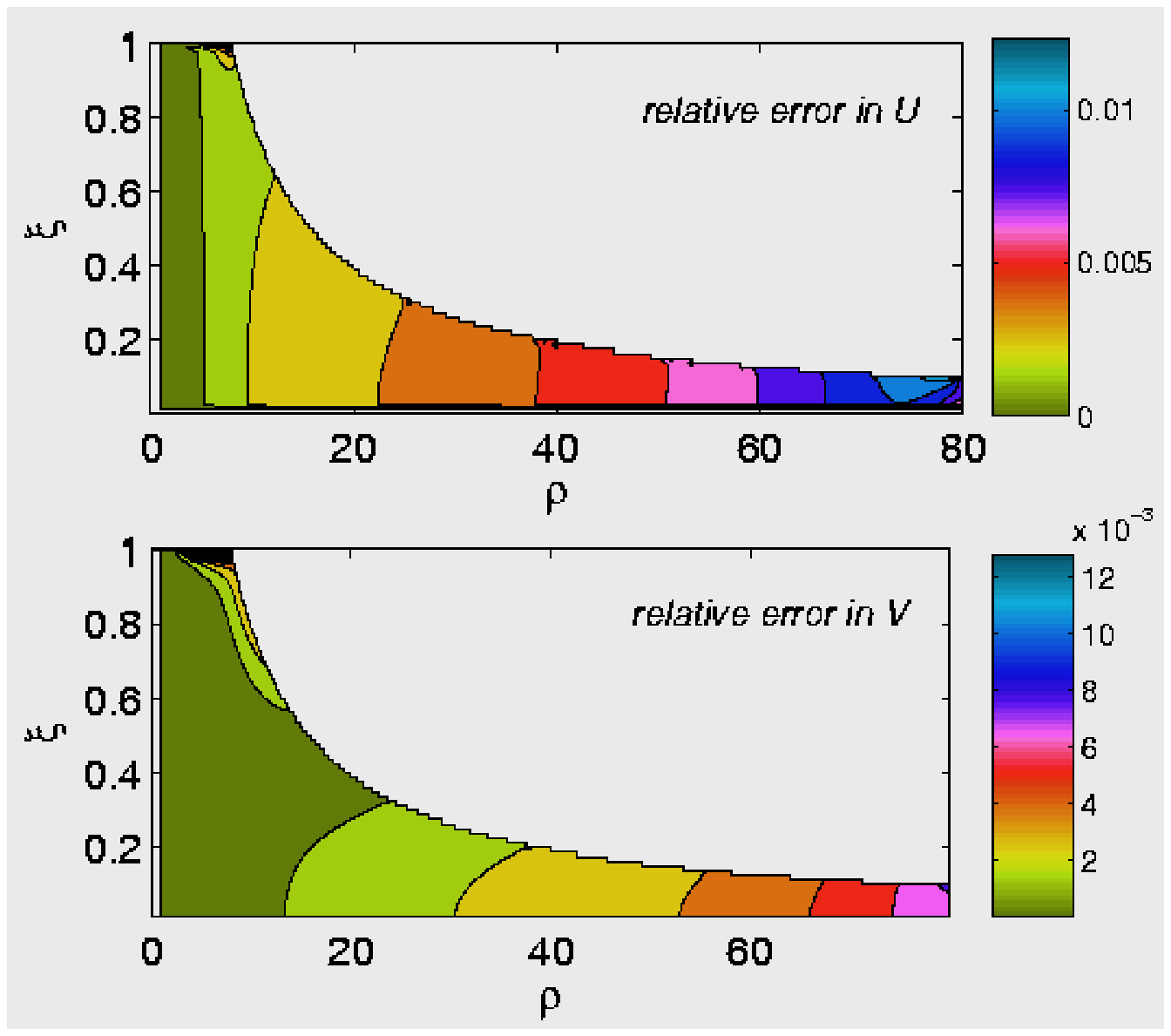}
\caption[]{ The relative errors in the constraints
  in the polar patch. These errors are small, being less then $2\%$,
  even though the absolute values of the constraints plotted in
  Fig.\ref{fig_Con_abs} are not. }
 \label{fig_Con_rel}
\end{figure}

The fast convergence, small errors and small constraint violation are
all attributes of the small $x$ runs. For certain $x$ the convergence
rate became noticeably slower and the errors became uncontrollable.
This was the $x$ when the ''problematic'' asymptotic $b$ started to
dominate.  Even though we may expect positive tension $\tau>0$, namely
$b<a/2$ in 5d $b$ reached $a/2$ and continued to grow for larger $x$.
This designated the maximal $x$ for which we could fully trust our results,
which is about $x \simeq 0.20$.

\sbsection{Applications of Smarr's formula}

Aside from numerical tests the Smarr formula (\ref{freeenergy})
provides an important theoretical constraint. We found that {\it
Smarr's formula is satisfied within $3-4 \%$}(!), see Fig.
\ref{fig_Smarr}. We find  a small  systematic error: when
evaluating $A_3 \kappa/(8 \pi a L)$, which should equal unity, we
find that the mean value of the numerical points is about $0.97$
with a mean spread of less then $2\%$. This shows that our
numerics produces systematically under-estimated values with a
relatively narrow spread. In addition, there is a slight increase
in accuracy when the asymptotic boundary is moved farther away. In
this case the center of the distribution moves toward unity,
though very slowly.

It is intriguing  that the Smarr formula is satisfied with good
accuracy for {\it all} our solutions, including those with a
problematic $b$. Even for $x>0.20$ when the ``b-dominance''
triggered convergence problems this highly nontrivial formula
continued to hold, see Fig.\ref{fig_Smarr}. Together with
Fig.\ref{fig_converg1} this suggests that despite the fact that we
could not determine the scalar charge accurately, the other three
measurables are determined with good accuracy.  Our working
assumption will be that these measurables are trustable, even for
$x>0.20$ up to the last convergent solution, at $x_2\simeq 0.25$,
though in this regime the assumption becomes somewhat speculative.
\begin{figure}[t!]
\centering
\noindent
\includegraphics[width=12cm]{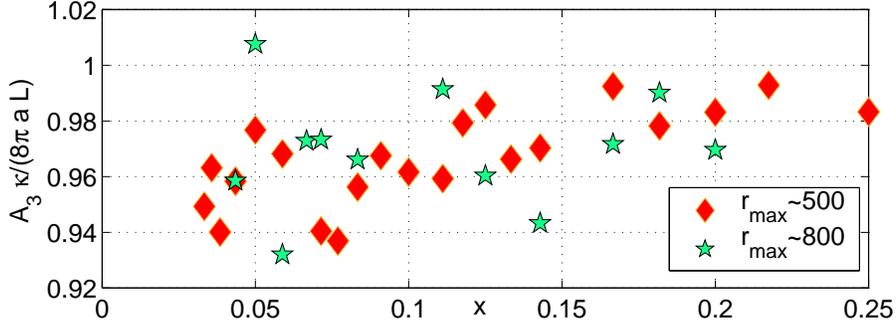}
\caption[]{ Testing the Smarr formula for two values of $r_{\rm
    max}$. The formula  predicts $A_3 \kappa/(8 \pi a L) =1$.
  The mean value for distribution of point designated by diamonds is
  $.967$ with standard deviation of $.017$. The same for the stars is
  $.97$ and $.02$ correspondingly.}
 \label{fig_Smarr}
\end{figure}
\begin{figure}[b!]
\centering
\noindent
\includegraphics[width=12cm]{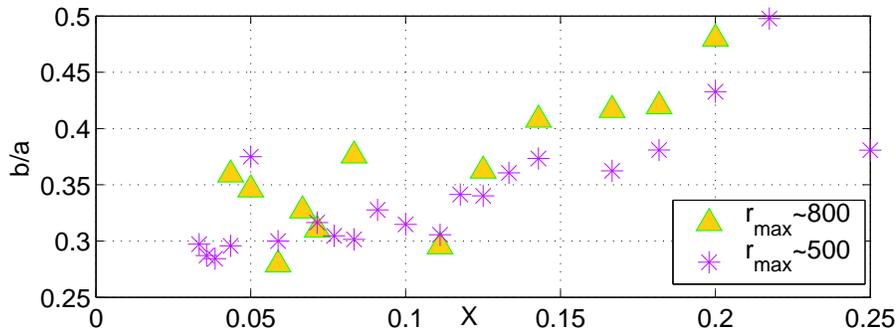}
\caption[]{ The ratio $b/a$ is expected to start at  $0.5$ for
  $x\ll 1$, and decrease for larger $x$. Since $b =0$ for uniform
  black string, one might speculate that $b/a \rightarrow 0$ as $x$
  increases. Here we see a different behavior.  This behavior is not
  reliable since $b$ is problematic (at $30\%$ level) even at small
  $x$.}
 \label{fig_b_a}
\end{figure}

Let us discuss one more aspect of the ``b-problem''. As we noted
before even for small $x$ $b$ did not really converge, see Fig.
\ref{fig_converg1}. In addition, assuming positive tension,
$\tau>0$, implies $b \leq a/2$, and the equality is obtained for
small black holes that have vanishing tension. Since $b=0$ for a
uniform string this suggests that for small $x$ $b/a$ should equal
$\sim 1/2$ and should decrease to zero as $x$ increases.  In
Fig.\ref{fig_b_a} we plot the ratio $b/a$. One observes that for
small $x$ values the points are distributed around $1/3$ rather
then around $1/2$. In accordance with our working assumption $a$
is calculated correctly, hence we estimate that the accuracy of
the b calculation constitutes $30 \%$, for small values of $x$. On
the other hand, we have seen that for  intermediate $0.08 \lesssim
x \lesssim 0.15$, $b$ does converge well.  Hence, since the ratio
$b/a$ is expected to decrease as $x$ increases, the measured value
of $b/a \sim 0.3$ in this $x$ range  can be trusted.
%
\section{Results}
\label{sec_results}
%
One of the most important results of this work is that our numerical
solutions are the first strong evidence for the existence of static
black holes in a non maximally symmetric higher dimensional spacetime.
We constructed a family of solutions parametrized by $x$, defined in
(\ref{iks}). The horizon region of the solutions in this branch tend
to the 5d \Sch \ solution in the $x \rightarrow 0$ limit and become
unstable for $ x \rightarrow 0.25$. Even before that, at $ x \simeq
0.20$, numerical errors were not small anymore.  Our analysis is not
capable to settle with certainty what has destabilized the algorithm.
We tend to interpret this  as originating from a physical
tachyonic instability, though currently we are not able to determine
the exact location of the transition point.

\sbsection{The geometry of the black holes}

Let us examine the spacetime structure of a typical member of our
family of solutions, with $x\sim 1/7$.  In Fig.
\ref{fig_spacetime} we use contour plots to visualize the behavior
of the metric functions and gain some insight for the geometry.
\begin{figure}[t!]
\centering
\noindent
\includegraphics[width=12cm]{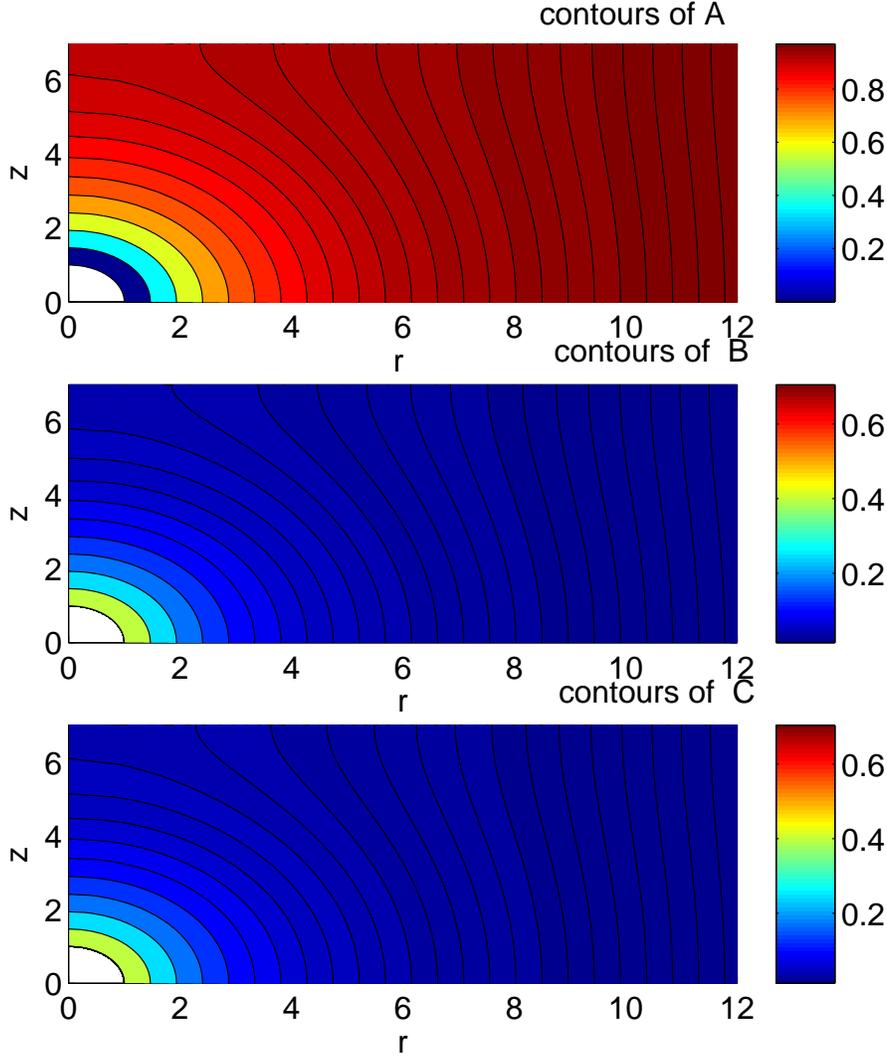}
\caption[]{ Contours of the metric functions for
  $x=1/7$.  The black hole is located at the left bottom corner. The
  vertical periodic direction marked by $z$.
  Note  the change of the topology of the contours.  Near the
  horizon they are spherical while far away the contours become
  cylindrical, indicating translation invariance along $z$ at infinity.}
\label{fig_spacetime}
\end{figure}
The function $A$ vanishes at the horizon and it approaches unity
asymptotically. The functions $B$ and $C$ decay smoothly from a finite
value at the horizon to zero at infinity. One observes as well that
the $z$-dependence disappears fast as one gets away from a black hole.
In addition, one learns that the contours intersect all the
boundaries at an angle of 90 degrees: the periodic at $z=L$, the reflecting at $z=0$,
and the $r=0$ axis. This shows that the
numerical scheme performs well at the boundaries.

The proper deformation or eccentricity of the horizon
(\ref{epsilon}) is depicted in Fig. \ref{fig_eps}.  Below $x\simeq
0.05$ the values that we obtain are randomly distributed around
zero with magnitudes of $10^{-6}$, hence we concluded that the
black hole is spherical to better then $10^{-6}$ at this regime.
The main tendency to be noted in the figure is that $\eps>0$. This
means that as $x$ grows the black hole becomes prolate along the
axis, tending to a string-like form, as one could expect
intuitively.  Another interesting feature is that the last black
hole that we have obtained (with $x\simeq 0.25$) is only slightly
deformed, with $\eps \simeq 6.5\cdot 10^{-3}$. Comparing with the
theoretical quantitative prediction (\ref{epsexpansion}), where
${8\over 3} \zeta(4) \simeq 2.89$, we see that the numerical
``experiment'' agrees exactly. Moreover, the axis intersection
value $-3.5\cdot10^{-5}$ agrees with the expected zero up to the
numerical errors, see Table \ref{table1}. We note that although
the agreement looks ``too good'', a fairly good agreement persists
also for fits on smaller neighborhoods of $x=0$.
\begin{figure}[t!]
\centering
\noindent
\includegraphics[width=10cm]{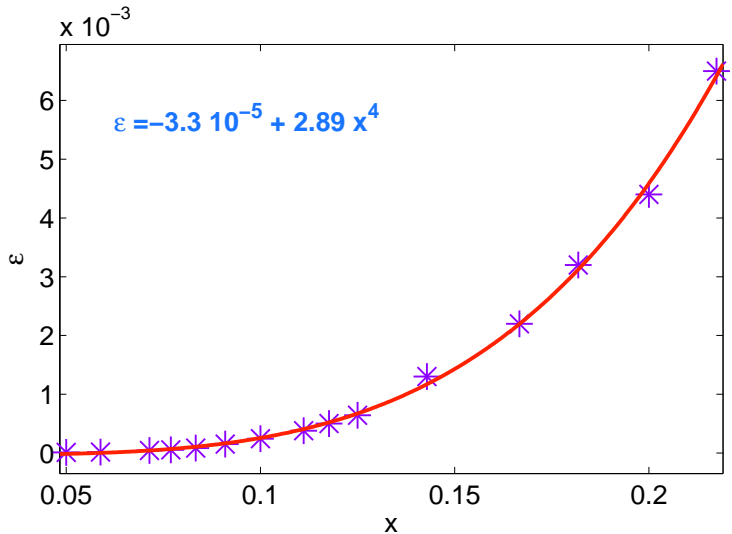}
\caption[]{Eccentricity or deformation of the horizon. It stays very
  small up to $x\simeq 0.25$, where our code becomes unstable. The
  coefficient of proportionality agrees exactly with (\ref{epsexpansion})!
  }
\label{fig_eps}
\end{figure}
\begin{figure}[h!]
\centering
\noindent
\includegraphics[width=8cm]{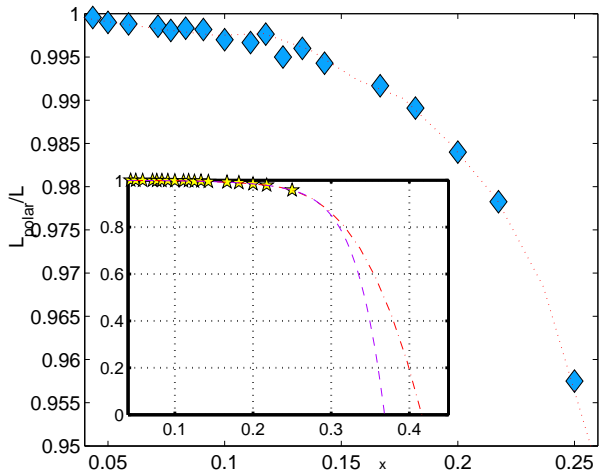}
\caption[]{The normalized inter-polar proper distance starts at 1 for
  small $x$ and decreases as $x$ grows. Note the surprisingly small rate of decrease. The insert contains a
  speculative {\it extrapolation}. If the latter is correct, the
  ``north'' and the ``south'' poles of the black hole will touch at
  $x\simeq 0.4$. The two extrapolation lines correspond to a spline
  (dash-dotted) and to a 8-degree polynomial (dashed). }
\label{fig_ell}
\end{figure}

The (normalized) inter-polar distance (\ref{polardistance1}) is
plotted in Fig. \ref{fig_ell}. One observes that $\ell$ decreases so
that the ``north'' and the ``south'' poles of the black hole approach
each other.  This decrease however is slow, such that just before the
last $x\simeq 0.25$, $\ell$ is still very far from vanishing.  This
suggests several possibilities: (1) Our numerical solution crashes due
to uncontrollable errors, and hence at $x\simeq 0.25$ there is not
really a tachyonic instability.  In this case the possibility that
$\ell$ will shrink to zero and the phase transition will be smooth is
not precluded by our analysis. (2) The point $x\simeq 0.25$ is in the
vicinity of the real phase transition and $\ell$ does not drop to
zero, signaling a non-smooth phase transition.  This possibility was
advocated in \cite{barak1}. It is interesting to note that such
features as the finiteness of $\ell$ and the smallness of the
eccentricity, just before the transition, were obtained in
\cite{sorpir} for momentary-static black hole solutions. The small $x$
behavior of $\ell$ is surprising. Neglecting the effect of the black
hole on the spacetime metric one would expect a linear decrease.
However, as the figure shows the decrease is much slower\footnote{It
  is hard to fit but seems quadric.}.  This phenomena is not
understood yet. It looks like as if the mass of the BH expands space
in such a way that compensates almost exactly for its size. This
effect can be called an ``Archimedes law for caged black holes''.

 In an insert in Fig. \ref{fig_ell}  we plotted a
possible extrapolation of the measured $\ell$. If this extrapolation
is correct then the poles will touch for $x\sim 0.4$.

\sbsection{Thermodynamical variables}

Our prime thermodynamical variables are depicted in Fig.
\ref{fig_4ThermoVars}.
\begin{figure}[t!]
\centering
\noindent
\includegraphics[width=14cm]{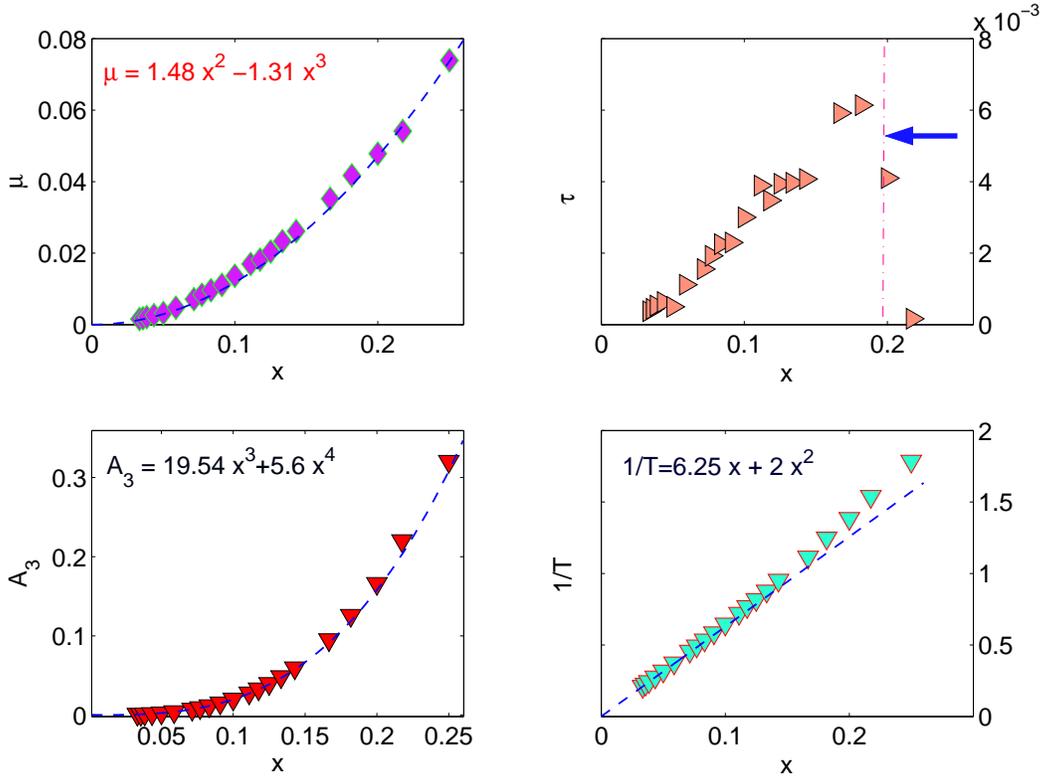}
\caption[]{ The dimensionless  mass, tension, 3-area and the
  temperature along the found branch. The dashed line designates the
  corresponding variables for the 5d \Sch \ solution.  The shown
  equations are an approximation for small $x$, the coefficients with
  the fitting  error are given in Table. \ref{table1}.  }
\label{fig_4ThermoVars}
\end{figure}
We see that in the small-$x$ limit all variables tend to their
\Sch \ values, designated by the thick dashed line. The
uncompactified \Sch \ solution appears to be a smooth limit of the
near horizon region of the caged black holes under discussion. One
notes that three of the four thermodynamical variables have a
smooth behavior all the way up to $x\simeq 0.25$, while the fourth
one (the tension) is somewhat less consistent and for
$x\gtrsim0.20$ its behavior is strange. This is in agreement with
our observations in section \ref{sec_tests}.  We have argued,
based on the success of the Smarr formula, that the three prime
measurables ${\cal A}_3,T,a$ are robust, while the fourth
variable, $b$, has an uncertainty of about $30\%$ based on
convergence problems and disagreement for small $x$. This rather
large error is presumably confined to $b$ due to an approximate
decoupling of equations in the asymptotic region, see appendix
\ref{appendix_asympt}. Moreover, it is this decoupling that allows
the success of Smarr's formula. Due to different relative weights
that the asymptotics $a$ and $b$ have in the mass and the tension
calculation, see (\ref{asymp_to_charges}), $\mu$ appears smooth in
Fig. \ref{fig_4ThermoVars}, while $\tau$ does not.

For small $x \ll 1$ one can consider a Taylor expansion of the
thermodynamical variables in powers of $x$.  We can attempt to
extract the expansion coefficients by fitting the  measurables with
polynomials.  We applied a fitting procedure for the first 8 to 10
solutions for which $x\leq 0.11 $.  The expansion coefficients
together with the numerical errors are given in  Table
\ref{table1}, as well as the expansion coefficients of
$A^{(\kappa)}$, defined in Eq.  (\ref{Akappa3}) and plotted in
Fig.\ref{fig_Akappa}.
\begin{figure}[t!]
\centering
\noindent
\includegraphics[width=9cm]{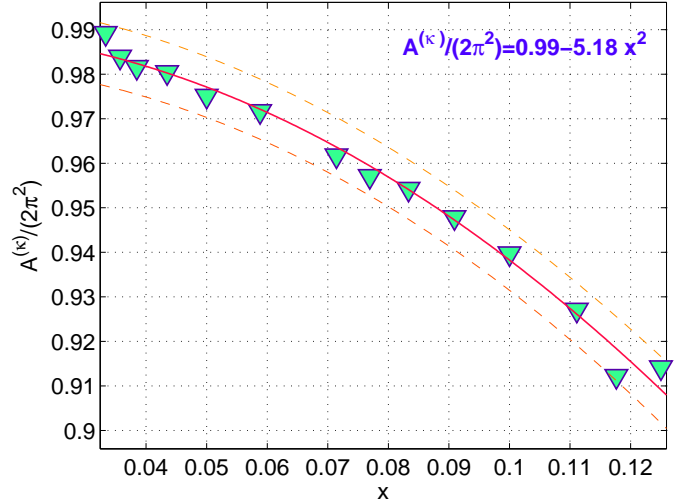}
\caption[]{ $A^{(\kappa)}/(2 \pi^2)$ for small $x$.
  The leading correction agrees well with (\ref{Akappa3expansion})! The
  confidence bounds are depicted.  No other Taylor coefficients could
  be extracted reliably.}
\label{fig_Akappa}
\end{figure}
For completeness, the corresponding data for $\eps$ is also added to
that table. Everywhere the numerical coefficients are
given with the estimated error. This is defined here by the variation
of the coefficients of the fitting functions, while allowing the fit
to vary within the $95\%$ confidence range, which is illustrated in
Fig.\ref{fig_Akappa}.
\begin{table}[b!]
\centering
\noindent
\begin{tabular}{|c|c|c|c|c|}\hline
&  \rm{Fitting Formula}& $f_1\pm \delta f_1$ & $f_2\pm \delta
f_2$& Theoretical  $f_1$  \\ \hline
$\mu$ &$f_1 x^2+f_2 x^3$  &$1.48\pm.07$ &$-1.3 \pm 0.6$& $3 \pi/8\simeq 1.18$ \\ \hline
${\cal A}_3$&$f_1 x^3+f_2 x^4$  &$ 19.54 \pm .08$ &$ 5.6\pm 1.0$& $2
\pi^2 \simeq 19.74$ \\ \hline
$T^{-1}$&$f_1 x+f_2 x^2$  & $6.25\pm .05 $& $2.0\pm .12 $ & $2\pi \simeq6.28$  \\ \hline
$A^{(\kappa)}/(2\pi^2)$ &$f_1+f_2 x^2$ &$ .99\pm.004   $& $-5.18\pm.4$ &$f_1=1,~f_2\simeq-4.93$\\ \hline
$\eps$ &$f_1+f_2 x^4$ &$ -(3.3\pm 4.8) \cdot 10^{-5}  $& $2.89\pm.06
$&$f_1=0, ~f_2\simeq  2.89$ \\ \hline
\end{tabular}
\caption[]{The numerically computed  expansion coefficients of
  the thermodynamical and geometric  variables. Theoretical
predictions for the leading terms are listed in the
last column. }
\label{table1}
\end{table}

We can compare now the expansion coefficients to the theoretical
predications that are summarized in section \ref{sec_theory}.  The
leading expansion coefficients, which are listed in the last
column of Table \ref{table1}, are confirmed with a good
confidence, aside for the mass, $\mu$, for which the numeric and
the theoretical numbers differ by about $20\%$. This is the
imprint of the poor $b$ behavior, since $b$ is a part of the
formula (\ref{asymp_to_charges}) for the mass. The higher order
corrections to $\eps$ and $A^{(\kappa)}$ match beautifully with 
 new theoretical results \cite{GorbonosKol,numericI}.  Other coefficients have a somewhat
greater uncertainty (sometimes tens of percents).  Since we do not
have a theoretical insight, and because of the large numerical
errors we regard them as tentative.

We conclude that in the small $x$ regime most of our measurables are
very robust. We argue as well, that the problematic $b$ affects
$\mu$, but it is even more influential in the tension calculation, see
Eqn.(\ref{asymp_to_charges}). The latter is depicted in Fig.
\ref{fig_4ThermoVars}. In this plot the demarcation line corresponds
to $x \simeq 0.20$.  In our numerical relaxation we observed a slow
down of convergence and a loss of accuracy when approaching $x\simeq
0.20$. This fact finds its vivid representation in the tension plot.
The points beyond the demarcation line drop suddenly and the tension
vanishes at $x\simeq 0.22$.\footnote{ We expect that the tension, much like the mass, is always
positive \cite{PositiveTraschen,PositiveSIT} so we regard this
behavior of $\tau$ as fictions resulting from the loss of accuracy in
the measurable  $b$.} 

Based on the success of the integrated first law our working
assumption is that the entropy, temperature and the numerical
asymptotic $a$ are robust not only for small $x$ but for all the
solutions, up to $x\simeq 0.25$, see the discussion in the end of
previous section \ref{sec_tests}. Moreover, one may observe that even
though the mass evaluation is very sensitive to $b$ the main
contributor is still $a$: $\mu = (a-b/2)/\hL$. Hence, as long as
$b<a/2$, which is the case here, see Fig.\ref{fig_b_a}, one may assume
that the mass calculation is accurate within $20-25\%$ limits. This
will be our second, more speculative, assumption.  Using it, one can
question what additional information can be extracted from out data.
Addressing this question we ask: is there

\sbsection{A phase transition?}

We interpret the instability of the numerics as a manifestation of a
real physical tachyonic instability, which slows down our scheme for
$x\gtrsim x_1\simeq0.20$, and which finally ruins it completely at
$x_2\simeq 0.25$.  Examining the (dimensionless) masses corresponding
to the above two $x$-values, we find that $\mu_1 \simeq 0.047$ and
$\mu_2 \simeq 0.074$ are not that far from the GL critical mass,
$\mu_{\rm GL} \simeq 0.070$. Since a priori, the various instability masses
in the system are expected to be of the same order, this coincidence
is rather suggestive. This is another evidence to our assumption that
we are approaching the real physical instability.
\begin{figure}[t]
\centering
\noindent
\includegraphics[width=10cm]{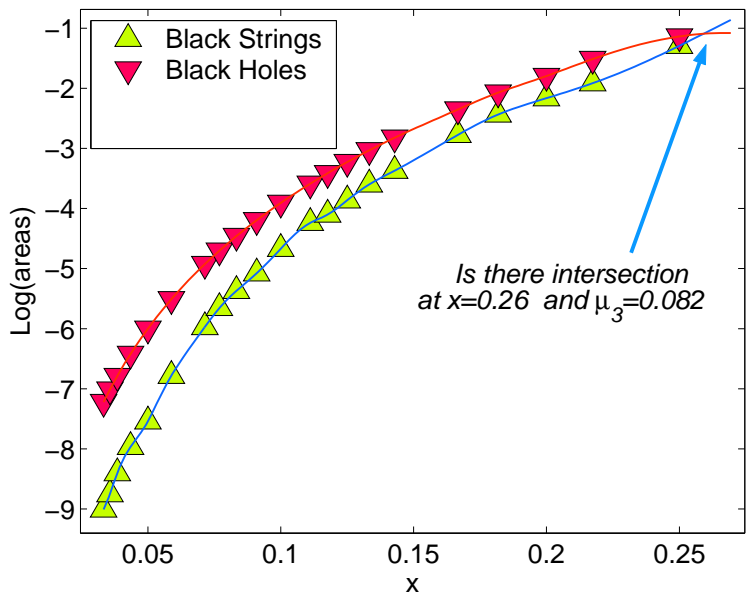}
\caption[]{ The logarithm of the dimensionless area of the black
  hole, ${\cal A}_3$, and black string, ${\cal A}_{BS}$, with the
  same mass. The continuous lines show extrapolation of the data to
  larger $x$ values, suggesting an intersection and a first order
  phase transition at $x\simeq 0.26$. }
\label{fig_Areas}
\end{figure}

The existence of a maximal mass designates perturbative, classical
instability. On the other hand, once the entropies of the two solutions
are equal for a given mass, a first order phase transition between the
two phases can take place. This transition will occur either by
quantum mechanical tunneling or by thermal fluctuations.  In Fig.
\ref{fig_Areas} we depict the logarithm of the dimensionless areas
${\cal A}$, for the two phases: the black hole and the black string,
where ${\cal A}_{BS}$, is computed for {\it the same mass},
$\mu=\mu(x)$.  Our data does not show a crossing of the areas.
However, a naive extrapolation of the data points, marked by the solid
lines in Fig.\ref{fig_Areas}, indicates an intersection just above our
last BH namely at $x\simeq 0.26$.  For this value the mass is $\mu_3
\simeq 0.082$, which is also of order of the critical GL mass.

Note that the extrapolated $\mu_3$ is slightly larger then our
instability mass, $\mu_2$, while the opposite is expected for a first
order phase transition. However, this inequality is not numerically
significant: due to the high degree of
uncertainties near the instability point we cannot estimate well the critical  mass $\mu_2$. In addition one cannot expect
that $\mu_3$ is evaluated accurately, since its value will depend
strongly on several (not very accurate) last points in
Fig.\ref{fig_Areas}.

It is important to determine whether $\mu_2 < \mu_{\rm GL}$. If so
there must be a third stable phase to which the black hole decays, for
instance the stable non-uniform string \cite{HM1}. However, we expect $\mu_2 > \mu_{\rm GL}$.

\section{Future directions}
\label{sec_discussion}

We conclude by pointing out some future directions

\begin{itemize}

\item {\it Higher dimensions $d>5$.}  In this paper we presented a 5d
  numerical implementation of the ideas outlined in our previous paper
  \cite{numericI}.  In principle, after some slight modifications the
  code can be applied to a higher dimensional problem.  While we
  expect that the instability, being physical, would still be present
  there, we hope to improve the accuracy of the critical mass
  estimation.

  We hope that the accuracy will improve in this case due to faster
  asymptotic decay.  In addition, as we described in section
  \ref{sec_tweaks} faster asymptotic decay implies smaller $r_{\rm
    max}$, hence smaller number of grid points and therefore faster
  operation of the code, which is now frustratingly long (1-2 days).

\item {\it Improving the performance of the method}. One can try to improve
  further the algorithm, by e.g. using the full multigrid technique
  that incorporates motion up and down grid numbers. In order to improve
  the accuracy near the axis we could use as an angular coordinate the
  angle $\chi$ instead of $\xi=\cos(\chi)$. The benefit is that the axis is
  approached faster in the $\chi$ coordinates, and so there are more grid
  points near it. The drawbacks (as we explained in the text) are that the
  coordinate singularity would be second order and the periodic
  boundary,$z=L$, would loose its particularly simple representation
  $\rho=L/\xi$.

\item{ \it Another choice of coordinates.}  The metric ansatz that we
  used here contains three functions. It would be interesting to try and implement the coordinates suggested in \cite{HO1} and substantiated in \cite{wis2,HO3}. In these  coordinates the  number of functions reduces to two, and moreover they interpolate smoothly between spherical coordinates in the horizon region and cylindrical ones asymptotically, thus eliminating the need for the two coordinate patches. 

\end{itemize}

\vspace{0.5cm} \noindent {\bf Acknowledgments}

It is a pleasure to thank J. Bahcall for suggesting this
collaboration. BK thanks T. Wiseman and D. Gorbonos for
discussions and collaboration on related issues.

The authors are supported in part by the Israeli Science
Foundation.

\appendix
\section{Equations of motion and boundary conditions  on a d-cylinder}
\label{Equations_appendix}

In this appendix we derive the equations of motion on a
d-dimensional cylinder, ${\mathbb R}^{d-2,1} \times \IS^1$ and the
corresponding boundary conditions

The most general ansatz which is time independent, time reversal
symmetric (``static'') and with axial  $SO(3)$ symmetry is
\beq \label{first_ansatz} ds^2=-\exp(2\hA) dt^2 + d\sigma^2(r,z)+
\exp(2\hC) d\Omega_{d-3}^{~2} \enq
where $z$ and $z+ \hL$ are identified , $~d\Omega_{d-3}^2$, is the $d-3$ sphere, $\hA,\hC$
are functions of $(r,z)$ and $d\sigma^2(r,z)$ is an arbitrary
metric in the $(r,z)$ plane.  Note that we dropped for now the
$r^2$ prefactor in front of $\exp(2\hC)$. Classically the problem
scales with $\hL$, and so we can set $\hL=2 \pi$ and it can always
be restored by dimensional analysis.

We need the expression for the Ricci scalar of a fibration. For
$ds^2=ds^2_X+ \exp(2 F(x))ds^2_Y$ one has
\beq
 R=R_X + \exp(-2F) R_Y - d_Y(d_Y+1) (\pa F)^2 - 2 d_Y \triangle(F)
\enq
where the Laplacian is given by $\triangle (F)=
\mbox{det}(g)^{-1/2} \pa_\mu (g^{\mu\nu}\mbox{det}(g)^{1/2}
\pa_\nu F)$, grad squared is $(\pa F)^2=g^{\mu\nu} ~\pa_\mu F
~\pa_\nu F$ and $d_X=\mbox{dim}(X), ~d_Y=\mbox{dim}(Y)$ (we
normalized the Ricci scalar by $R=R_{ijij}$ so that $R_{{\bf
S}^d}= d(d-1)$).

The d-dimensional Ricci scalar is
\bea R_d&=&R_2 - 2(\pa \hA)^2-2 \triangle \hA +(d-3)\,(d-4)\,
e^{-2\hC}\non &-& (d-3)\,(d-2)\,(\pa \hC)^2 - 2\,(d-3)\, \triangle
\hC - 2(d-3) (\pa \hA)(\pa \hC) \ena
 While deriving this formula
one has to be careful to consider the $\hA, ~\hC$ fibration
step-wise, and thereby get the cross term $-2(d-3) (\pa \hA)(\pa
\hC)$ in addition to $-2 \triangle(\hA+(d-3) \hC)$.

The gravitational action is \beq
 I={\Omega_{d-3}  \over 16 \pi G_N} \int dt dV_2 ~e^{\hA+(d-3)\, \hC} R_d ,
\enq
where $G_N$ is the d-dimensional Newton constant, $dV_2=
\det_{(r,z)}\, dr dz $,  $\Omega_{d-3}$ is the area of the unit
$d-3$ sphere, and from now on we will drop the prefactor
${\Omega_{d-3} \over 16 \pi G_N} \int dt$.

After integrating by parts we get
\beq I=\int dV_2 ~e^{\hA+(d-3) \hC}[R_2 + (d-3)(d-4) e^{-2\hC} +
(d-3)(d-4) (\pa \hC)^2 + 2(d-3)(\pa \hA)(\pa \hC)]. \enq
Now we need to fix the metric ansatz in the 2d $(r,z)$ space.  One
can use diffeomorphism invariance to put it in the conformal form
\beq
 ds^2=-\exp(2\hA) dt^2 + \exp(2\hB)(dr^2+dz^2)+
\exp(2\hC) d\Omega^2. \enq
The formula for the Ricci scalar of a conformally transformed
metric  $g^{ab}=\exp(2\hB) \tilde{g}^{ab}$ reads \cite{wald}
\beq
 R_d=e^{-2\hB}[\tilde{R}_d-2 (d-1) \pa_{ii} \hB- (d-1)(d-2) \pa_i \hB \pa_i \hB].
\enq
Hence in 2d with the conformally flat metric we get
\beq
 R_2 = -2 e^{-2\hB} \pa_{ii} \hB .
\enq
The action (without second derivatives after integration by parts)
is
\bea I= \int dr ~dz [  (d-3)(d&-&4) e^{\hA+2\hB+(d-5)\hC} +
e^{\hA+(d-3)\hC}( 2\,\pa_i \hA \pa_i \hB + 2(d-3) \pa_i\hA \pa_i
\hC \non &+& 2(d-3) \pa_i \hB \pa_i \hC + (d-3)(d-4)\pa_i \hC
\pa_i \hC)].  \ena

The equations of motion are
\bea
\hA : && \pa_{ii}(\hB+(d-3) \hC) + {(d-3)(d-2)\over2} (\pa_ i \hC)^2- {(d-3)(d-4)\over2}e^{2\hB-2\hC}=0 \\
\hB : && \pa_{ii}(\hA+(d-3)\hC) + (\pa_ i (\hA+(d-3)\hC))^2- (d-3)(d-4) e^{2\hB-2\hC}=0, \\
\hC : && \pa_{ii}( \hA+\hB+(d-4)\hC)+(\pa_ i \hA)^2 + (d-4)(\pa_ i
\hA)(\pa_ i \hC) \non && + {(d-3)(d-4)\over2}(\pa_ i
\hC)^2-{(d-4)(d-5)\over2}e^{2\hB-2\hC}=0. \ena
where the grad squared is defined here {\it without} a metric
factor $(\pa_i \hA)^2=(\pa_i \hA)(\pa_i \hA)$.

We find it convenient to redefine $$\hB \rightarrow B,~~ e^{\hat
C} \rightarrow e^C r ~~ {\rm
  and} ~~ e^{\hat A} \rightarrow A.$$
Solving for $\triangle(A),\, \triangle(B),\, \triangle(C)$ we
obtain the equations of motion, which in $\rz$ coordinates are
\bea \label{rz_d_A} &&\triangle A +(d-3) {\pa_r A\over r}
+(d-3)(\pa_r A \pa_r C +\pa_z A
\pa_z C) =0, \\
\label{rz_d_B} &&\triangle B-{(d-3)(d-4)\over2}\pa_r C
\left({2\over r} +\pa_r
  C\right)-{(d-3)\over2}\pa_z C\left((d-4)\,\pa_z C+2 {\pa_z A\over
    A}\right)\non
&&-(d-3) {\pa_r A\over A}\left({1\over r} +\pa_r C \right)
-{(d-4)(d-3)\over2}{1-e^{2 B -2 C}\over r^2}=0 , \\
\label{rz_d_C} &&\triangle C+(d-3)\pa_r C \left({2\over r} +\pa_r
C\right)+\pa_z C\left((d-3)\,\pa_z C+ {\pa_z A\over A}\right)\non
&&+{\pa_r A\over
  A}\left({1\over r} +\pa_r C \right) +(d-4){1-e^{2 B -2 C}\over
  r^2}=0 , \ena where $\triangle := \pa_r^2+\pa_z^2$.
These equations can be transformed to polar coordinates,
$\{\rho,\chi\}$, where we have
\bea \label{rhochi_d_A} \triangle A &+&(d-3) {\pa_\rho A\over
\rho} +(d-3)\pa_\rho A \pa_\rho C
+{(d-3)\pa_\chi A \over \rho^2 } (\pa_\chi C+{\rm ctg}(\chi)) =0, \\
\label{rhochi_d_B} \triangle B&-&{(d-3)(d-4)\over2}\left[\pa_\rho
C \left({2\over \rho} +\pa_\rho
  C\right)+{\pa_\chi C\over\rho^2}\left(\pa_\chi C+2
  {\rm{ctg}}(\chi)\right)\right] -\non
&-&{(d-3)\over A} \left[ \pa_\rho A\left(\pa_\rho C+{1 \over \rho
      }\right)+ {\pa_\chi A \over \rho^2} (\pa_\chi C +{\rm
    ctg}(\chi))\right] -\\
&-&{(d-4)(d-3)\over2}{1-e^{2 B -2 C}\over \rho^2 \sin(\chi)^2}=0 ,
\non \label{rhochi_d_C} \triangle C &+&(d-3)\left[\pa_\rho C
\left({2\over \rho} +\pa_\rho
  C\right)+{\pa_\chi C\over\rho^2}\left(\pa_\chi C+2
  {\rm{ctg}}(\chi)\right)\right] +\non
&+&{1\over A} \left[ \pa_\rho A\left(\pa_\rho C+{1 \over \rho
      }\right)+ {\pa_\chi A \over \rho^2} (\pa_\chi C +{\rm
    ctg}(\chi))\right] +(d-4){1-e^{2 B -2 C}\over \rho^2
  \sin(\chi)^2}=0 . \
\ena
 Here $\triangle :=
 \pa_\rho^2+(1/\rho)\pa_\rho+(1/\rho^2)\pa_\chi^2$ is the flat
 Laplacian in the polar coordinates.

 These are elliptic equations, and as such they are subject to
 {\it boundary conditions}. One can see that the boundary conditions are
 basically dimension independent. The only difference will appear at
 the boundary at infinity, because of the different fall-off rate of the
 functions.  Asymptotic flatness singles out natural radial coordinate
 at infinity, $r \sim r_{\rm \Sch}$. In \cite{numericI} we found for $d>5$
\bea 1-A &=& {a \over r^{d-4}} + O({1
   \over r^{d-3}}), \non B &=& {b \over r^{d-4}} + O({1 \over
   r^{d-3}}), \non C &=& {c \over r} + O({1 \over r^{d-4}}) \label{defab}
 \ena
 The feature to be noted is that C is the slowest decaying function,
 with the same rate for all dimensions $d>5$. The case
 $d=5$ is somewhat special since  $C$ decays as   $\log(r)/r$ at the
 leading order.

\section{Asymptotic behavior.}
\label{appendix_asympt}

At infinity  the equations become one-dimensional, as all
z-dependence is washed out  exponentially fast.  Defining for
convenience $\zeta := \log(r)$, and retaining in
(\ref{rz_d_A}-\ref{rz_d_C}) only the $r$-dependent terms, we
obtain:
\bea \label{rz_d_A1}
&& A'' +(d-4)A' +(d-3)A'C'  =0, \\
\label{rz_d_B1} &&B''-B'-{(d-3)(d-4)\over2}C'\left(2 +C'\right)
-(d-3) {A'\over A}\left(1 + C' \right) \non
&&-{(d-4)(d-3)\over2}\left(1-e^{2 B -2 C}\right)=0,~~~ \\
\label{rz_d_C1} &&C''-C'+(d-3) C' \left(2 + C'\right)+{A'\over
  A}\left(1+C' \right) +(d-4)\left(1-e^{2 B -2 C}\right)=0 , \ena
where $(~)':= d/d{\zeta}$.

Linearizing these equations around flat space, $A-1=B=C=0$, one
obtains
 \bea
\label{rz_d_A2}
&& A'' +(d-4)A'   =0, \\
\label{rz_d_B2}
&&B''-B'-(d-3)(d-4)C'+(d-4)(d-3)(B -C) -(d-3)A'=0, \\
\label{rz_d_C2} &&C''+2(d-{7\over2}) C' +2(d-4)(C-B)+A'=0 . \ena

One observes that the first equation {\it decouples}\footnote{This
  effective decoupling ensures that the Smarr formula is satisfied very
  accurately even though $b$ is somewhat less accurate.} from the other
two and it can be solved to yield $A \simeq 1-a e^{-(d-4)\zeta} =
1-a/r^{d-4}$.  In 5d the solutions to the other two equations are
$B=b/r$ and $C=c_5\log (r) /r$ together with $c_5=-a+2\, b$
\cite{numericI}, which can be checked by explicit substitution.

For a stability analysis let us look for a solution of  form $ B,C
\sim B_0,C_0\cdot exp(ikr)$ to the other two equations. A
tachyonic mode would be one such that ${\rm Im}(k)<0$.  After
substitution in the homogeneous equations one gets the algebraic
equations
\beq  \left[
  \begin{array}{cc}
    -k^2 -ik +(d-3)(d-4)  & -ik(d-3)(d-4) -(d-3)(d-4) \\
    -2(d-4)   & -k^2 +2(d-{7\over2})ik +2(d-4) \\
   \end{array} \right] \,
 \left[ \begin{array}{c} B_0 \\ C_0 \\ \end{array} \right]=0.
\enq
which have a unique solution when the determinant of the matrix
vanishes \beq i \,\left( -5 + d \right) \, \left( -4 + d \right)
\,k - \left( 11 - 7\,d + d^2 \right) \,k^2 - 2\,i \,\left( -4 + d
\right) \,k^3 + k^4 =0. \enq The solutions of this equation are
\beq \label{inf_modes} k_1=0, ~~~ k_2=i(d-5), ~~~ k_3=i (d-4), ~~~
k_4=i .  \enq
Since for all the modes, ${\rm Im} (k)>0$ there is no tachyon.
Note that there is a massless mode $k_1=0$ (in 5d $k_2$ is
massless as well) that corresponds to the choice $C=B$
asymptotically. This massless mode can, in principle, become the
unstable one due to non-linear corrections or numerical errors,
but we have no explicit indication for this.


\end{document}